\documentclass[aip,graphicx]{revtex4-1}
\usepackage{graphicx}
\usepackage{epstopdf, epsfig}
\usepackage{mathptmx}
\usepackage{subcaption}
\usepackage{bm}
\usepackage{amsfonts}
\usepackage{float}
\usepackage{hyperref}
\hypersetup{
	colorlinks=true,
	linkcolor=black,
	filecolor=blue,      
	urlcolor=black,
	citecolor=black
}
\draft 

\begin{document}


\title{Role of corner flow separation in unsteady dynamics of hypersonic flow over a double wedge geometry} 



\author{Gaurav Kumar, Ashoke De}
\email[]{ashoke@iitk.ac.in}
\affiliation{Department of Aerospace Engineering, Indian Institute of Technology, Kanpur, UP, India - 208016}


\date{\today}

\begin{abstract}
This study investigates the origin and sustenance of self-induced oscillations of shock structures in a hypersonic flow over a double wedge configuration. Previously, various researchers have considered the double wedge flow configuration for inviscid flow with variations of different inflow as well as geometric parameters such as inflow Mach number ($M_\infty$), wedge angles, and wedge lengths. Few recently published articles reveal an unsteady flow physics involved with the hypersonic viscous flow for double wedge configuration with large second wedge angles. However, the reason for such self-sustained flow oscillations is not completely clear. The present work seeks out to investigate the origin of such oscillations in a low enthalpy hypersonic flow with different aft-wedge angles and wedge length ratios. In the current study, viscous flow over a double wedge at $M_\infty$ = 7 and fore-wedge angle of $30^\circ$ is considered. An improved version of rhoCentralFoam solver in OpenFOAM is used to investigate the double wedge flow over different grid resolutions in the separation region and shock-shock interaction region. This study corroborates the observation from the previous literature with an improvement in the range of parameters which results in a self-sustained periodic oscillation. The present study also suggests that the unsteadiness becomes possible when the incidence shock is in the proximity of the aft-wedge expansion corner as a consequence of different wedge length ratios ($L_1/L_2$) or aft-wedge angles ($\theta_2$). Flow can still be steady at a large aft-wedge angle if the incidence shock is far ahead of the aft-wedge expansion corner. 
\end{abstract}

\pacs{}

\maketitle 

\section{\label{sec:intro}Introduction}
Double wedge configuration for studying hypersonic flows over solid surfaces is an ingenious idealization. It is geometrically simple yet produces a diversity of complex flow physics. Double wedge configuration is useful for understanding the high thermo-mechanical loads generated near corners contained between the control surfaces and the fuselage of hypersonic vehicles. High thermo-mechanical loads on the vehicle can cause loss of control as well as spoil the structural integrity of the vehicle. The catastrophic failure of the X-15 aircraft is an apt example of this problem in the research community. Also, a hypersonic flow intake is designed with a similar double ramp configuration to provide compression of inflow for the operation of scramjet engines. A large unsteadiness in the flow or significant separation can reduce the efficiency of the engine; moreover, it causes instabilities\cite{devaraj2020experimental,sekar2020unsteady}.

Although the experimental study of hypersonic flows is a challenging task due to the scarcity of observation techniques as well as the inordinate cost of assembling and running such a facility, a continuous effort has been made towards understanding such high thermal and mechanical loads in hypersonic flows. The earliest account is from ref.\cite{edney1968anomalous}, where shock-shock interaction mechanisms have been categorized into six types, and their roles are identified in the generation of high peak pressure and peak heat-transfer rates near shock impingement point on the model. This categorization has been widely acknowledged in analyzing and understanding shock-shock interactions in all the further experimental as well as numerical studies. Olejniczak\cite{olejniczak1996high} first studied the flow over double wedge experimentally and numerically to validate the non-equilibrium chemistry models used in Computational Fluid Dynamics (CFD). They found a discrepancy between heat transfer rates obtained from CFD compared to experiments and attributed it to the deficiencies of the chemistry models; however, the flow was very dependent on the size of the separation zone predicted. Later, Olejniczak \cite{olejniczak1997numerical} further investigated inviscid shock-shock interactions and transition between different types of interactions depending on the second wedge angle in a double wedge configuration exposed to supersonic as well as hypersonic flows. The study performed for a diatomic perfect gas ($\gamma$ = 1.4), Mach number ($M_\infty$) = 9, fore-wedge angle ($\theta_1$) = $15^\circ$, fore to aft-wedge length ratio ($L_1/L_2$) = 1 and aft-wedge angle variation from $35^\circ$ to $60^\circ$ revealed that the inviscid shock-shock interaction mechanism transitions from type VI $\rightarrow$ V $\rightarrow$ IV $\rightarrow$ IVr. This interaction mechanism results in an under-expanded supersonic jet along the surface and causes high oscillations in pressure along the wall surface. Ben-dor\cite{ben2003self} in their study showed a self-induced oscillation of the shock system in the range of $42^\circ < \theta_2 < 43^\circ$ with the same flow configuration as described in ref.\cite{olejniczak1997numerical} with $L_1/L_2$ = 2.  Later, a high-resolution numerical study was performed in ref.\cite{hu2008numerical}, where $M_\infty$ = 7 and 9 were simulated over a double wedge with $\theta_1=15^\circ$ and $L_1/L_2$=2. They have considered the variation of caloric properties of the gas with temperature, but the flow was still modeled as inviscid. A difference in $\theta_2$ was observed compared to the study of ref.\cite{ben2003self}, where self-induced oscillations occur, and it was demonstrated that the difference was due to the difference of gas model used and not due to the numerical resolutions or schemes. It was also conjectured that a viscous dissipation might damp down such self-sustained oscillations of the shock system due to a change in flow topology. Still, no detailed viscous computations were carried out due to numerical difficulties.

In ref.\cite{swantek2012heat}, a range of new experimental data was put forward regarding shock wave and separation region interaction on a double wedge configuration and the coupling between fluid mechanics and thermochemistry of hypersonic flow. This work has attracted a portion of the CFD community because of the availability of heat transfer rate and flow visualization through schlieren data for a high fore-wedge angle ($\theta_1$)=$30^\circ$ as well as an aft-wedge angle ($\theta_2$)=$55^\circ$ where the flow separation at the corner is significant. A large number of research groups participated in a study to predict the flow features of ref.\cite{swantek2012heat}, numerically, in ref.\cite{knight2017assessment} for both low enthalpy as well as high enthalpy flows where thermochemistry plays an important role. Various programs have achieved different degrees of accuracy for predicting overall flow features and wall heat transfer rates. Still, most of the simulations show issues in matching the separated flow region with the experiment. The double wedge case in ref.\cite{swantek2012heat} was first numerically investigated in detail by Komieves et al. in ref.\cite{komives2014numerical}. It was observed that a numerically first-order accurate solution could predict the wall heat flux in the separation region and shock impingement regions very well compared to the experiment but failed to predict it right in the attached flow region, bringing caution while interpreting a solution method that is numerically less accurate. Further investigation with second-order numerical methods revealed more detailed features, and the flow was seen to reach a low-frequency periodic state. 3D simulations showed differences with the 2D simulation results, and span-wise variations in the flow was seen. However, it should also be noted that the wedge has a low span-wise to fore-wedge length ratio ($L_z/L_1$) = 2, which results in a large contribution of edge effect from the geometry. Overall, none of the computational results were found to be in complete agreement with the experiment.  Some other published research also points towards the effect of side walls on the three-dimensionality of the separation region for high speed flows.\cite{funderburk2016experimental,huang2016evolution} Durna et al.\cite{durna2016shock} also investigated the detailed flow physics involved in the low enthalpy case of ref.\cite{swantek2012heat} with Nitrogen ($N_2$) as a working fluid where the effect of thermochemistry is negligible with the complete viscous flow simulation. Effect of variation of aft-wedge angle ($45^\circ<\theta_2<60^\circ$) was also considered to find out the change in shock-shock interaction mechanism. In recent work, Durna et al.\cite{durna2019time} further showed that in a long term unsteady simulation of this flow configuration, the flow never achieves a steady-state rather it settles down in a periodic self-sustained oscillation of the complete shock system. A similar observation was also made in ref.\cite{knight2017assessment}, contrary to the experiments in ref.\cite{swantek2012heat}, which was conducted for a very small time duration. In ref.\cite{durna2020effects}, Durna et al. further investigated the effects of flow three-dimensionality in the separated flow region at a large aft-wedge angle ($\theta_2 = 55^\circ$). The study reveals 3-D flow structures are influencing overall shock-structures and thermo-mechanical loading, which is mainly due to low span-wise to a fore-wedge length of the wedge ($L_z/L_1$=2). Still, there is a large discrepancy between experimental and numerical wall heat transfer rate prediction. A more recent numerical work in ref.\cite{reinert2020simulations} further showed that the flow is completely in the continuum regime, and other observations about separation region and unsteadiness from the previous pieces of literature were corroborated.

The investigations conducted by Durna et al. in \cite{durna2016shock,knight2017assessment,durna2019time,durna2020effects} reveal valuable insight in the physics related to the shock-shock and shock-boundary layer/separation region on the double wedge configuration. These simulations are performed using a solver called \textit{rhoCentralFoam}\cite{greenshields2010implementation} in OpenFOAM\cite{greenshields2015openfoam}, which is an open-source C++ based CFD platform. This solver has been extensively used for various kinds of supersonic and hypersonic flow simulations in the CFD community for a long time\cite{soni2018role,soni2018investigation}. This solver is second-order accurate in space and only first-order accurate in time. This deficiency produces many issues in simulating unsteady hypersonic flows. Authors have found out that the sequential time-integration of inviscid fluxes of conservation equations used in this solver has made this solver inherently unstable for time integration at a high CFL number ($\approx 1.0$). This kind of restriction is well known to the OpenFOAM CFD community, and a relatively lower CFL number ($\approx 0.2$) is recommended for the CFD simulations. This, still, can produce small oscillations and instabilities in the solution and make a long-duration simulation expensive due to the choice of a very small time step. Moreover, the use of a first-order time integration method also contributes to the above problem as well as adds excess numerical dissipation in unsteady simulation, which has not been of importance when the flow had to be studied only when in steady-state. Considering these limitations of {\it rhoCentralFoam} solver for a very long duration unsteady computation of hypersonic flow over a double wedge configuration, authors have developed an improved algorithm for \textit{rhoCentralFoam} solver in OpenFOAM framework which time integrates all conservation equations simultaneously removing the instabilities from \textit{rhoCentralFoam}. It is stable at CFL$\approx1.0$ for unsteady flow simulations. This improvement is further enhanced by the implementation of a $3^{rd}$ order TVD Runge-Kutta time integration method. A brief description of this new solver is given in sec. \ref{sec:math}.

Despite a large number of studies available on this topic, there is a large variation between the numerical predictions best summarized in ref.\cite{knight2017assessment} and actual experimental observations in ref.\cite{swantek2012heat}. Authors think these discrepancies can be attributed to two reasons. The first reason points towards the extremely short duration of the experiment in which the flow was not fully developed, especially the corner flow separation region. Due to the difference in the time-averaging window, a discrepancy in the wall heat flux can be seen. The second reason is the numerical inaccuracies, as pointed out in ref.\cite{komives2014numerical}. This manuscript provides more accurate computational validation of the experimental data along with the improved observation of unsteady flow over the double wedge configuration. The unsteadiness is shown to go away for the exact geometry of ref.\cite{swantek2012heat} with grid refinement, and a parameter regarding geometric configuration for shock reflection is identified, which results in a self-sustained oscillating shock system.

A brief description of the mathematical methods used in the current simulations is provided in sec. \ref{sec:math} and a few validation cases relevant to the resolution of current flow physics are shown in appendix \ref{Validation}. Computational setup for the double wedge flow is described in sec. \ref{sec:setup}. A detailed and systematics grid refinement study is performed in sec. \ref{sec:grid comparison} and the flow physics obtained from the computation is discussed in detail in sec. \ref{sec:result}. 

\section{\label{sec:math}Mathematical method}
For compressible flow simulation, a set of conservation equations of mass, momentum, and energy as given by eq. (\ref{conservationEqn}) is solved. Here $\vec{W} = [\rho\quad \rho \vec{u}\quad \rho E]^T$ represents mass density, linear momentum, and total energy of the control volume. 

\begin{equation}
\frac{\partial}{\partial t}\int_{\Omega}\vec{W} d\Omega + \oint_{\partial\Omega}(\vec{F_c} - \vec{F_\nu}) dS = 0
\label{conservationEqn}
\end{equation}

Here, $\vec{F_c}$ and $\vec{F_\nu}$ are convective and diffusive fluxes, respectively. The expressions for these terms are given in eq. (\ref{fluxRepresentation}). $\hat{n}$ represents unit normal vector to the faces of the control volume. 

\begin{equation}
\vec{F}_c = \left[
\begin{array}{c}
\rho(\vec{u}.\hat{n}) \\
\rho\vec{u}(\vec{u}.\hat{n})+p \hat{n}  \\
(\rho E + p)(\vec{u}.\hat{n}) \\
\end{array}  \right]\quad
\vec{F}_\nu = \left[
\begin{array}{c}
0 \\
\overline{\overline{\tau}}^T.\hat{n}  \\
(\overline{\overline{\tau}}^T.\hat{n}).\vec{u} + \kappa(\nabla T).\hat{n} \\
\end{array}  \right]
\label{fluxRepresentation}
\end{equation}

E is defined as $E = e+ \frac{1}{2}|\vec{u}|^2$ where e is the internal energy of the control volume. $e$ relates to temperature $T$ as $e(T)=e(T_0)+\int_{T_0}^{T}C_v(T)dT$ and variation of $C_v(T)$ with temperature is estimated using JANAF table\cite{chase1998nist}. $\overline{\overline{\tau}}= \mu[(\nabla \vec{u}) + (\nabla \vec{u})^T]  -\frac{2\mu}{3} tr(\nabla \vec{u})$ is the viscous stress tensor and $\mu$ and $\kappa$ are dynamic viscosity and thermal conductivity coefficient. In addition to eqs. (\ref{conservationEqn}) and (\ref{fluxRepresentation}), the state equation for an ideal gas is used. Sutherland relation accounts for the variation of $\mu$ and $k$ with temperature.

To solve the above set of governing equations, a new, improved algorithm for {\it rhoCentralFoam} is developed in OpenFOAM framework\cite{greenshields2015openfoam}. This solution method deploys an operator-splitting method that splits the inviscid and viscous part of the solution. The inviscid part of the solution is time-integrated explicitly using the third-order TVD Runge-Kutta method in conjunction with the flux-vector splitting method described in ref.\cite{greenshields2010implementation} with the K-T (Kurganov-Tadmor) method from ref.\cite{kurganov2000new}. The viscous term is discretized using $2^{nd}$ order central scheme, and the viscous corrector part of the solution is computed using an implicit temporal discretization of the governing equation.
\begin{figure}[h!]
	\centering
	\includegraphics[width=\linewidth, trim={0 2cm 0 2cm}, clip]{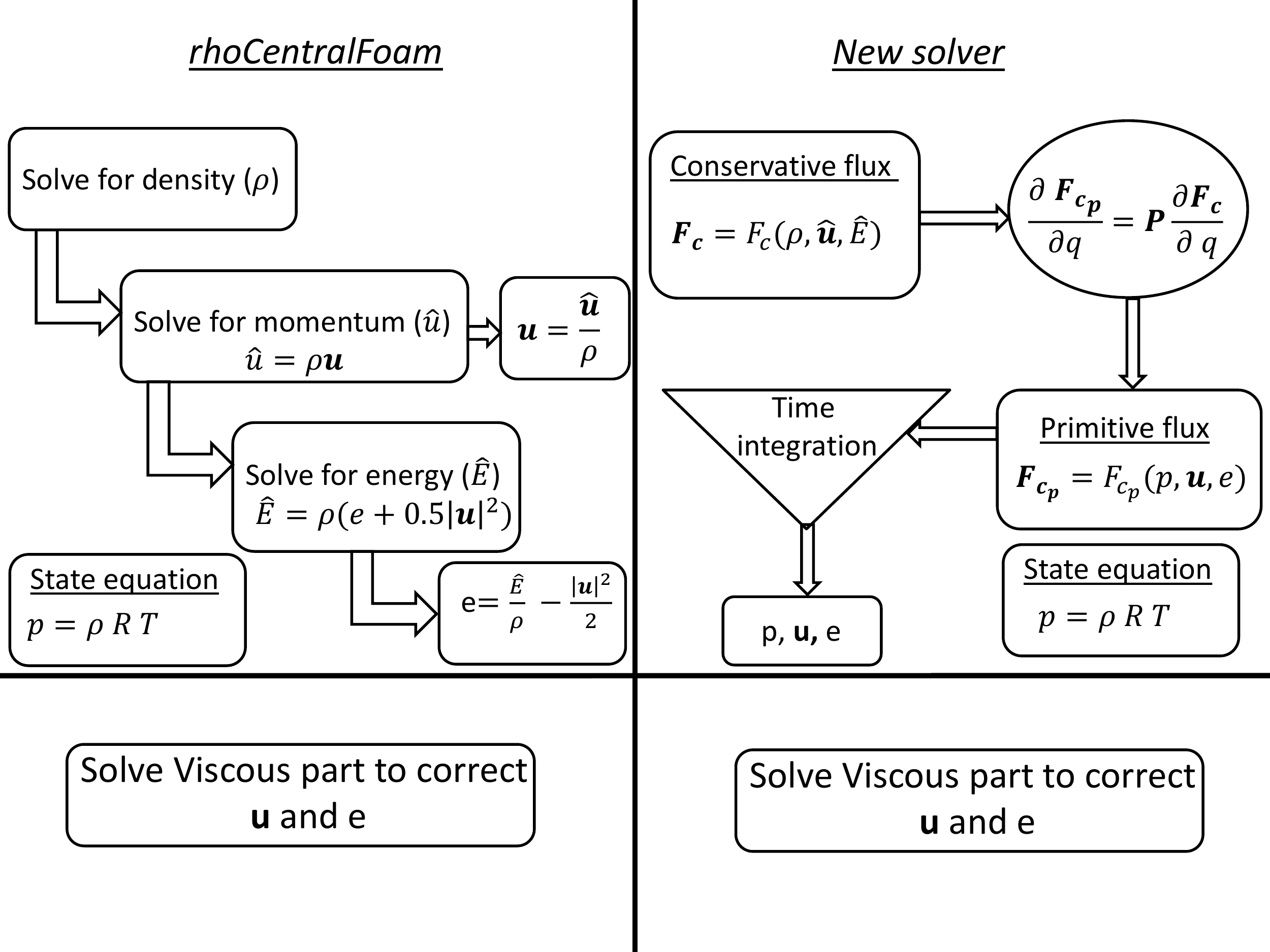}  
	\caption{Comparison between solution algorithm for the \textit{rhoCentralFoam} (left) and it's improved version (right)}
	\label{fig:algorithm}
\end{figure}
The main difference in the new solver compared to the previous solver \textit{rhoCentralFoam}, apart from the implementation of the high order time integration method, is time integration of inviscid flux. In \textit{rhoCentralFoam}, convective fluxes for mass, momentum, and energy are time-integrated sequentially, which causes instability in the computational method due to the use of a lagging field value from the previous time integration. To avoid this issue, the new solver time-integrates all three convective fluxes simultaneously for a single time step (or stage of Runge-Kutta integration method) so that no lag in the field value from previous time is created. For this, the field $\vec{W}$ is transformed in $\vec{W_p} = [p \quad \vec{u}\quad e]$ using a transformation matrix. Details of such a transformation matrix can be found in ref.\cite{blazek2015computational}. This difference in the algorithm is graphically shown in fig. \ref{fig:algorithm}. The accuracy of this new implementation has been verified for various engineering problems, and only a few test cases of interest are shown in appendix \ref{Validation} for validation. 

\section{\label{sec:setup}Computational setup}
\begin{figure}[h!]
	\centering
	\includegraphics[width=\linewidth]{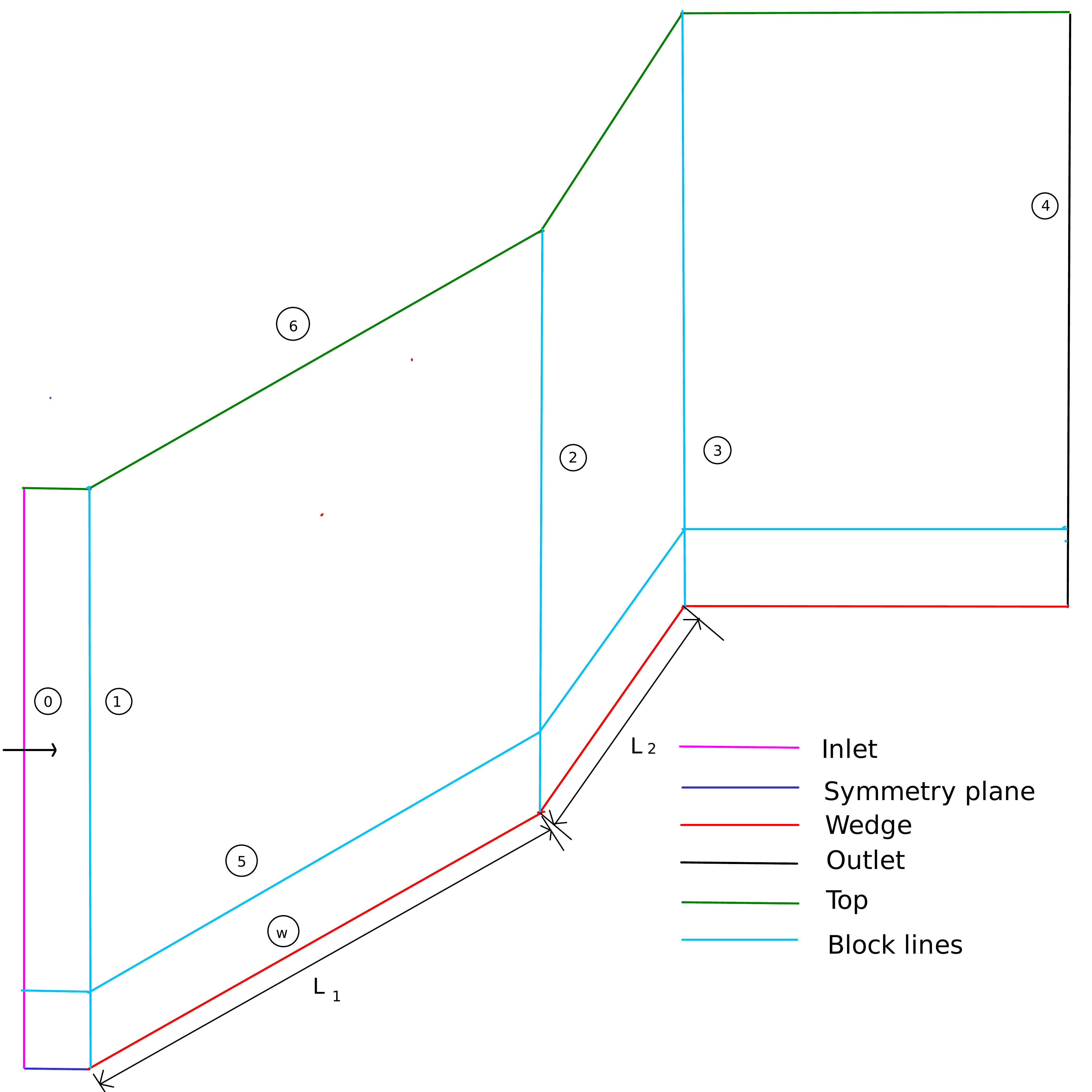}  
	\caption{Domain for double wedge configuration}
	\label{fig:dwDomain}
\end{figure}

Figure \ref{fig:dwDomain} illustrates the domain used for the computation of double wedge hypersonic flow. The fore-wedge length ($L_1$) is 50.8 mm (2 in) long, and aft-wedge length is $L_1$/2 based on the experimental setup of ref. \cite{swantek2012heat}. 
Inlet is placed at a distance $L_1/8$ upstream from the leading edge (LE) in order to remove the effect of enforced inlet boundary condition near the LE. A plane horizontal wall of length 0.75$L_1$ is used downstream of the second wedge in order to remove any effect of outlet to the flow over the wedges. For the same reason, the top boundary is also kept at a distance of 1.5$L_1$ to avoid any flow obstruction due to partial reflection from the outflow boundaries. For this simulation, the fore-wedge angle ($\theta_1$) is fixed at $30^\circ$, and the aft-wedge angle ($\theta_2$) is considered to be $55^\circ$ based on experiments in ref.\cite{swantek2012heat}.

A fixed inflow condition is provided at the inlet where velocity ($U_\infty$) = 1972 m/s, pressure ($P_\infty$) = 391 Pa, and temperature ($T_\infty$) = 191 K based on the experimental setup in ref.\cite{swantek2012heat}. This results in a freestream Mach no. ($M_\infty$) = 7.0. A symmetry boundary condition is provided at the symmetry plane, and the top and the right outlet are assigned wave transmissive boundary condition. Both the wedges and the downstream plane is considered isothermal no-slip viscous wall with surface temperature ($T_\infty$) = 300 K in accordance with the experimental setup in ref.\cite{swantek2012heat}. The flow in the internal domain is initialized with the inflow velocity ($U_\infty$), pressure ($P_\infty$) and temperature ($T_\infty$). Gaseous Nitrogen (N$_2$) is considered as the working fluid, and the JANAF tables\cite{chase1998nist} are used for calculating the caloric properties of the fluid at different temperature values. From the experiment\cite{swantek2012heat} and previous numerical computations\cite{durna2016shock,durna2019time,durna2020effects,knight2017assessment} it has been established that the flow is laminar. Hence, only a laminar flow simulation is considered here.

\section{Grid independence study\label{sec:grid comparison}}
In numerical simulations, truncation error could be accumulated either through insufficient grid size or large time-integration steps, or low order numerical schemes. Since a $3^{rd}$ order TVD Runge-Kutta time integration method with a second-order spatial accuracy is used, which can provide a stable solution at CFL number =1, all the simulations are performed at CFL=1. This results in time-step size in the range of 1 to 3 ns for all the simulations considered here, which is much smaller than the physical time scale. 

For a grid independence study, two sets of studies need to be considered. The first study involves sufficient resolution of gradients in the inviscid shock-shock interaction region and vortices in the separated flow region. The second study focuses on a sufficient wall resolution so that correct velocity and temperature gradients are predicted at the wall.

\subsection{Study 1: Resolution of gradients away from the wall}
A structured grid is generated within the domain, details of which are given in table \ref{tab:dwGrid}. Grid blocking information is shown using lines named "Block lines" in fig. \ref{fig:dwDomain}. Block line numbered "5" runs parallel to the wedge surface along with the domain at a height $0.15L_1$ such that on the aft-wedge it passes through the shock-shock intersection region, and on the fore-wedge it confines the separation region with the wall. In the first grid independence study, the grid is refined only between the aft-wedge compression corner and expansion corner. The details of variation in horizontal and vertical size of grid cells are shown in Table \ref{tab:dwGrid}.
\begin{table}[h!]
	\centering
	\begin{tabular}{|l |l |l |l |l |l |l |l |l |l |l |}\hline
		Grid & $\Delta x_0/L_1$ & $\Delta x_1/L_1$ & $\Delta x_2/L_1$ & $\Delta x_3/L_1$ & $\Delta x_4/L_1$ & $\Delta y_w/L_1$ & $\Delta y_1/L_1$ & $\Delta y_2/L_1$ & No. of cells & $U_\infty\overline{\Delta t}/L_1$\\ \hline
		Coarse & 0.01 & 0.01 & 0.01 & 0.01 & 0.025 & $10^{-4}$ & 0.01 & 0.04 & 91524 & $1.12\times10^{-4}$\\
		Medium & 0.01 & 0.01 & 0.005 & 0.005 & 0.025 & $10^{-4}$ & 0.005 & 0.04 & $1.99\times10^5$ & $7.92\times10^{-5}$ \\
		Fine & 0.01 & 0.01 & 0.001 & 0.001 & 0.025 & $10^{-4}$ & 0.001 & 0.04 & $1.27\times10^6$ & $4.98\times10^{-5}$ \\
		Finer & 0.01 & 0.01 & $5\times10^{-4}$ &  $5\times10^{-4}$ & 0.025 & $10^{-4}$ & $5\times10^{-4}$ & 0.04 & $2.95\times10^6$ & $4.6\times10^{-5}$ \\ \hline
	\end{tabular}
	\captionof{table}{Grid details for the first grid comparison study. $\Delta x_i$ represents horizontal cell length at vertical block line "i" and $\Delta y_j$ represents vertical cell length at horizontal block line "j" in fig. \ref{fig:dwDomain}.}
	\label{tab:dwGrid}
\end{table}
In the aft-wedge grid refinement, stream-wise grid stretching is avoided  so that the truncation error locally reduces quadratically with grid refinement, and $\Delta y_1 \approx \Delta x_1$ is chosen so that the local grid aspect ratio is close to 1. The boundary layer is resolved in the region 0.15$L_1$ above the wedges such that the first cell height above the wedges (w) is $10^{-4}L_1$. While refining the grid in the internal domain, grid sizes at the domain boundaries are kept constant, as shown in Table \ref{tab:dwGrid}. All the variations in the grid sizes between block lines are matched using a geometric grid stretching.

\begin{figure}[h!]
	\centering
	\includegraphics[width=\linewidth, trim={0 0 0 12cm}, clip]{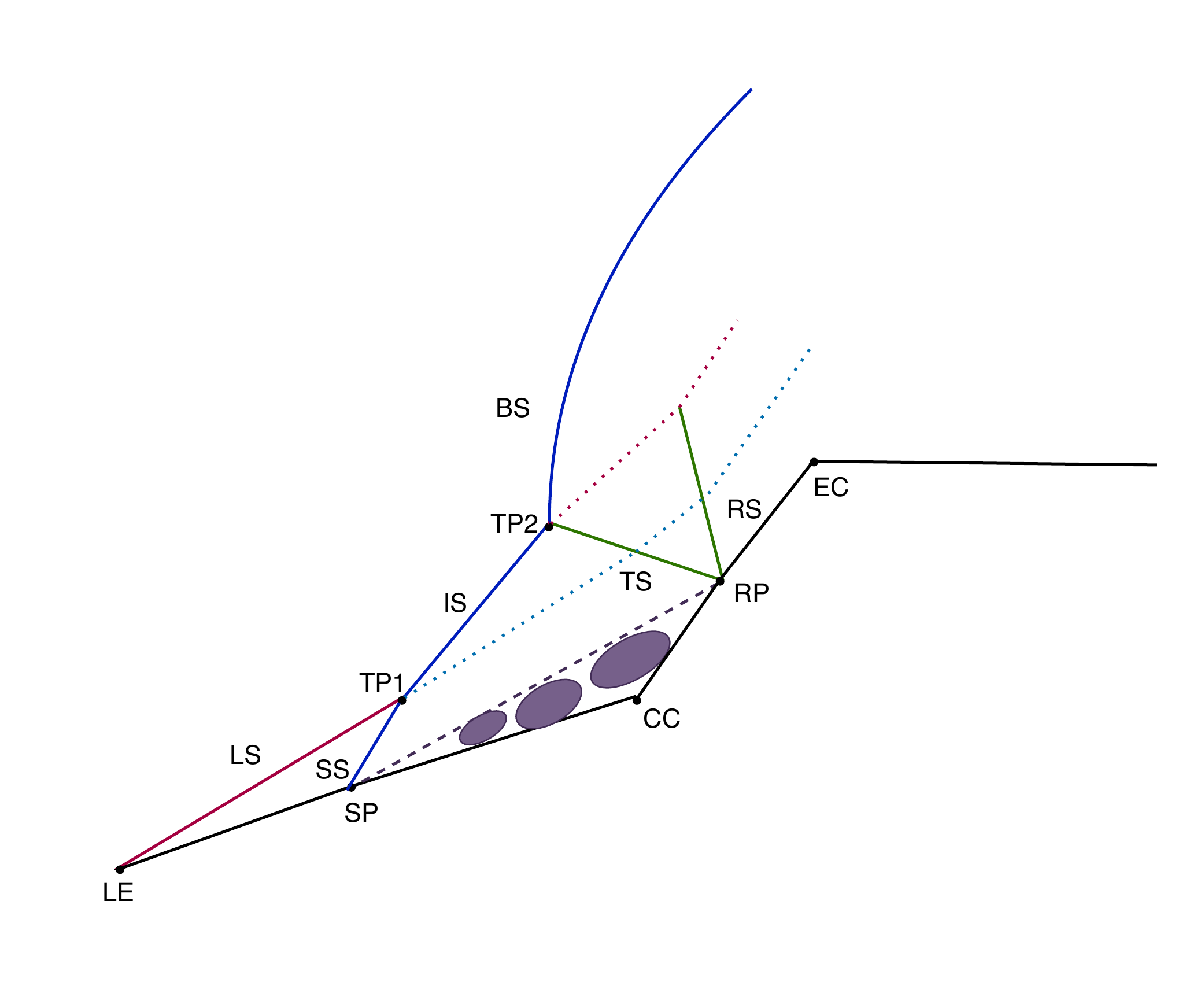}  
	\caption{A schematic of shock-shock interaction and shock reflection pattern with small separation region. Nomenclature: LE (Leading edge), CC (Compression corner), EC (Expansion corner), SP (Separation point), RP (Reattachment point), TP1 (First tripple point), TP2 (Second tripple point), LS(Leading edge shock), SS (Separation shock), IS (Intermediate shock), BS (Bow shock), RS (Reflected shock), TS (Transmitted shock). Dashed lines represent the shear layer.}
	\label{fig:sch260us}
\end{figure}

\begin{figure}[h!]
	\centering
	\includegraphics[width=\linewidth]{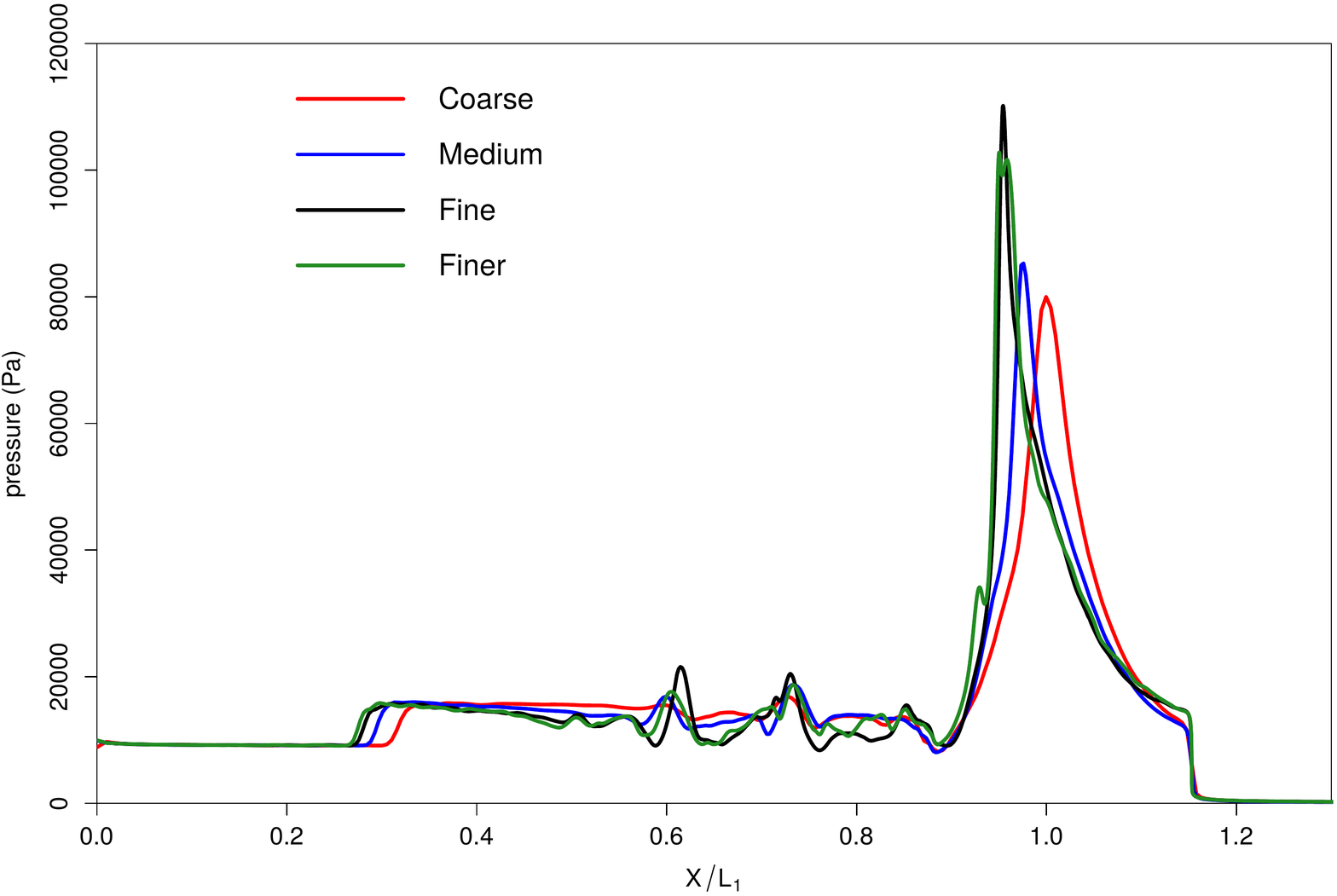}  
	\caption{Instantaneous wall pressure distribution with $\theta_2=55^\circ$ and $L_1/L_2=2$ at t = 260 $\mu s$}
	\label{fig:pressure260usStudy1}
\end{figure}

Figure \ref{fig:30_55_2_260us} shows a comparison of flow structure using the grids from Table \ref{tab:dwGrid} at t = 260 $\mu s$. Qualitatively all four grids provide very similar flow structures such as shock-shock interaction pattern and separation regions. The leading edge shock (LS)  and the Intermediate shock (IS) after LE shock interacts with the separation shock (SS) are closely comparable to one seen in schlieren photograph from Swantek's experiment\cite{swantek2012heat}. The variation in separation point location on the fore-wedge and the transmitted shock impingement location on the aft-wedge is also small among the simulations using four grid resolutions. But the major difference is seen in the resolution of the strength of the shear layer and size of vortices formed inside the separation region. Figure \ref{fig:sch260us} shows a schematic of the shock-shock interaction and reflection of transmitted shock from the wedge. This schematic provides an overall view of the interaction and reflection pattern omitting the complex details of the separation region. Figure \ref{fig:pressure260usStudy1} shows the distribution of wall pressure along the wedge length at t = 260 $\mu s$. The location of separation point (SP) on the fore-wedge region and shock impingement point/reattachment point (RP) on the aft-wedge region is shifted upstream when resolving grid from Coarse to the medium where cell sizes are halved.
\begin{figure}[h!]
	\centering
	\begin{subfigure}[t]{\linewidth}
		\centering
		\includegraphics[width=0.55\linewidth]{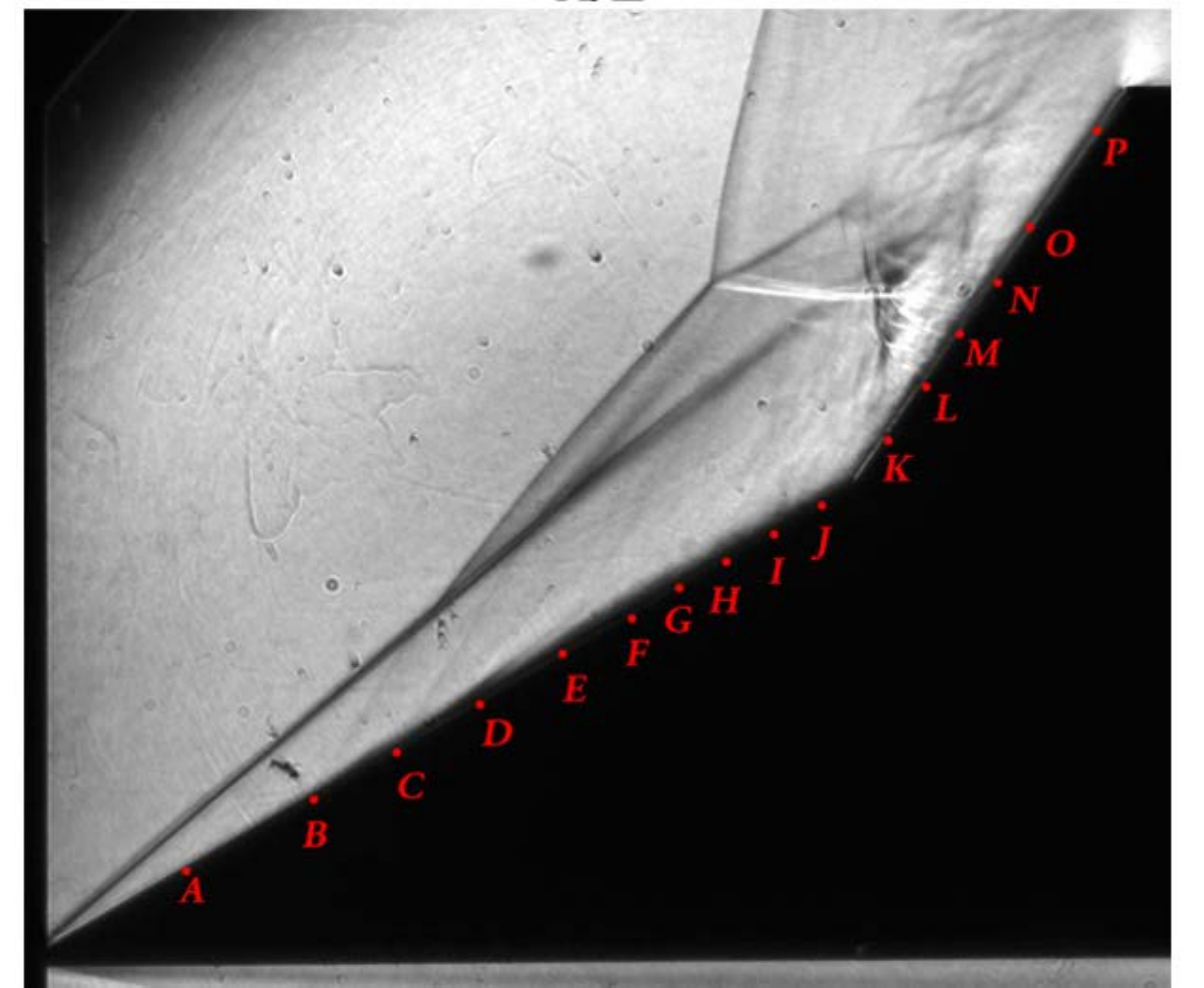}  
		\caption{}
		\label{fig:30_55_2_swantek_260us}
	\end{subfigure}
\end{figure}
\begin{figure}[h!]\ContinuedFloat
	\begin{subfigure}[t]{0.495\linewidth}
		\centering
		\includegraphics[width=\linewidth, trim={0 0 30cm 9cm}, clip]{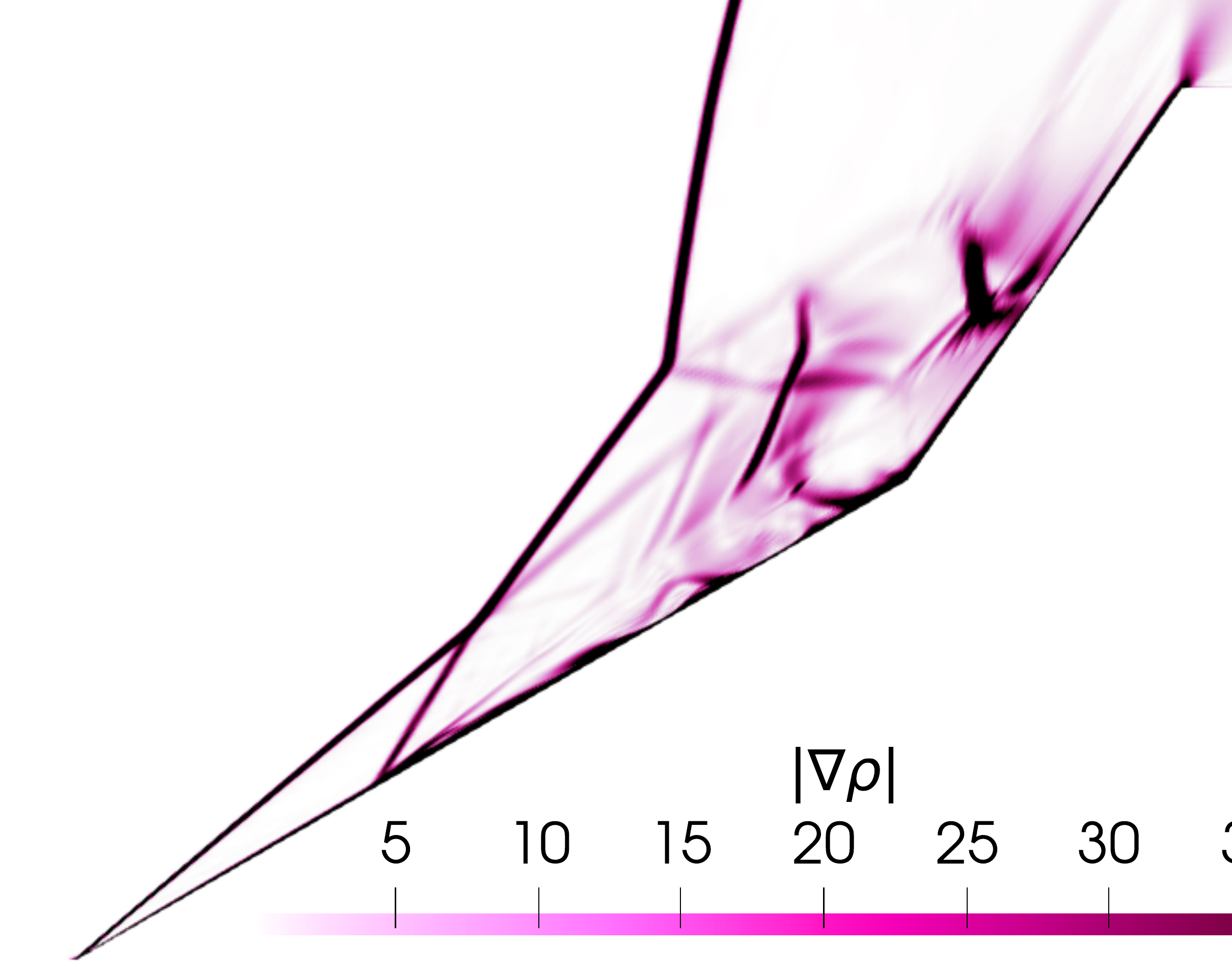}  
		\caption{}
		\label{fig:30_55_2_Coarse_260us}
	\end{subfigure}
	\begin{subfigure}[t]{0.495\linewidth}
		\centering
		\includegraphics[width=\linewidth, trim={0 0 30cm 9cm}, clip]{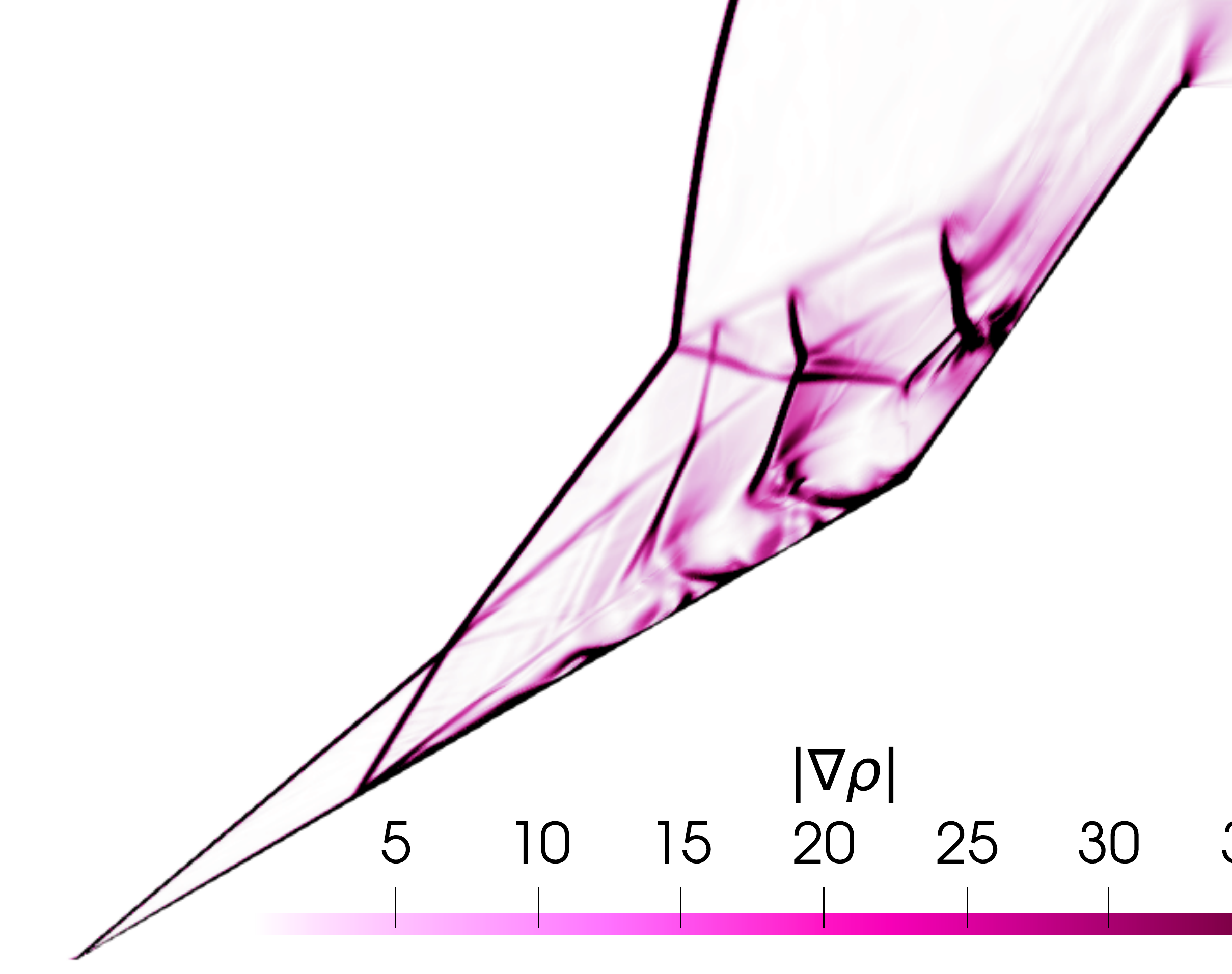}  
		\caption{}
		\label{fig:30_55_2_Medium_260us}
	\end{subfigure}
	\begin{subfigure}[t]{0.495\linewidth}
		\centering
		\includegraphics[width=\linewidth, trim={0 0 30cm 9cm}, clip]{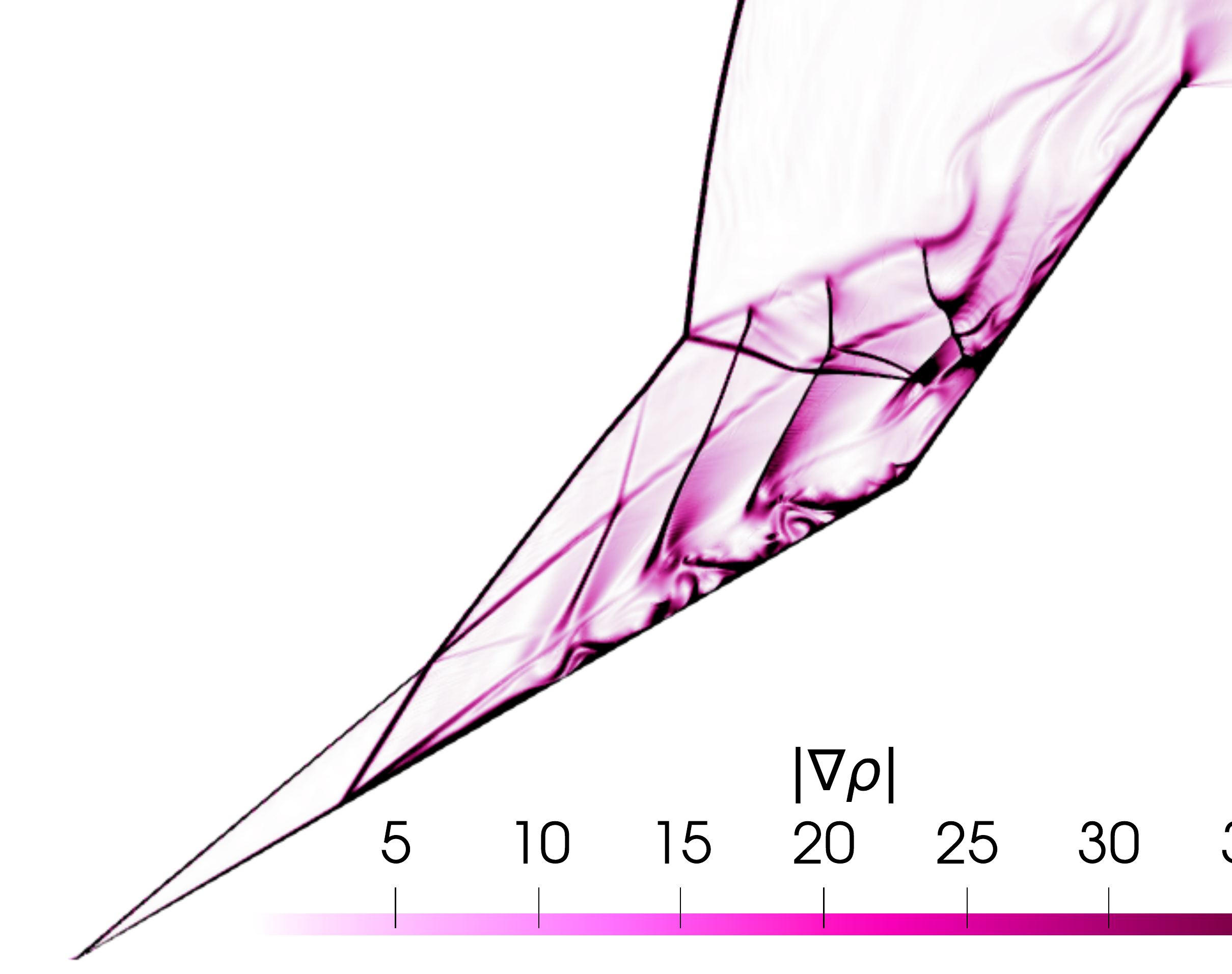}  
		\caption{}
		\label{fig:30_55_2_Fine_260us}
	\end{subfigure}
	\begin{subfigure}[t]{0.495\linewidth}
		\centering
		\includegraphics[width=\linewidth, trim={0 0 30cm 9cm}, clip]{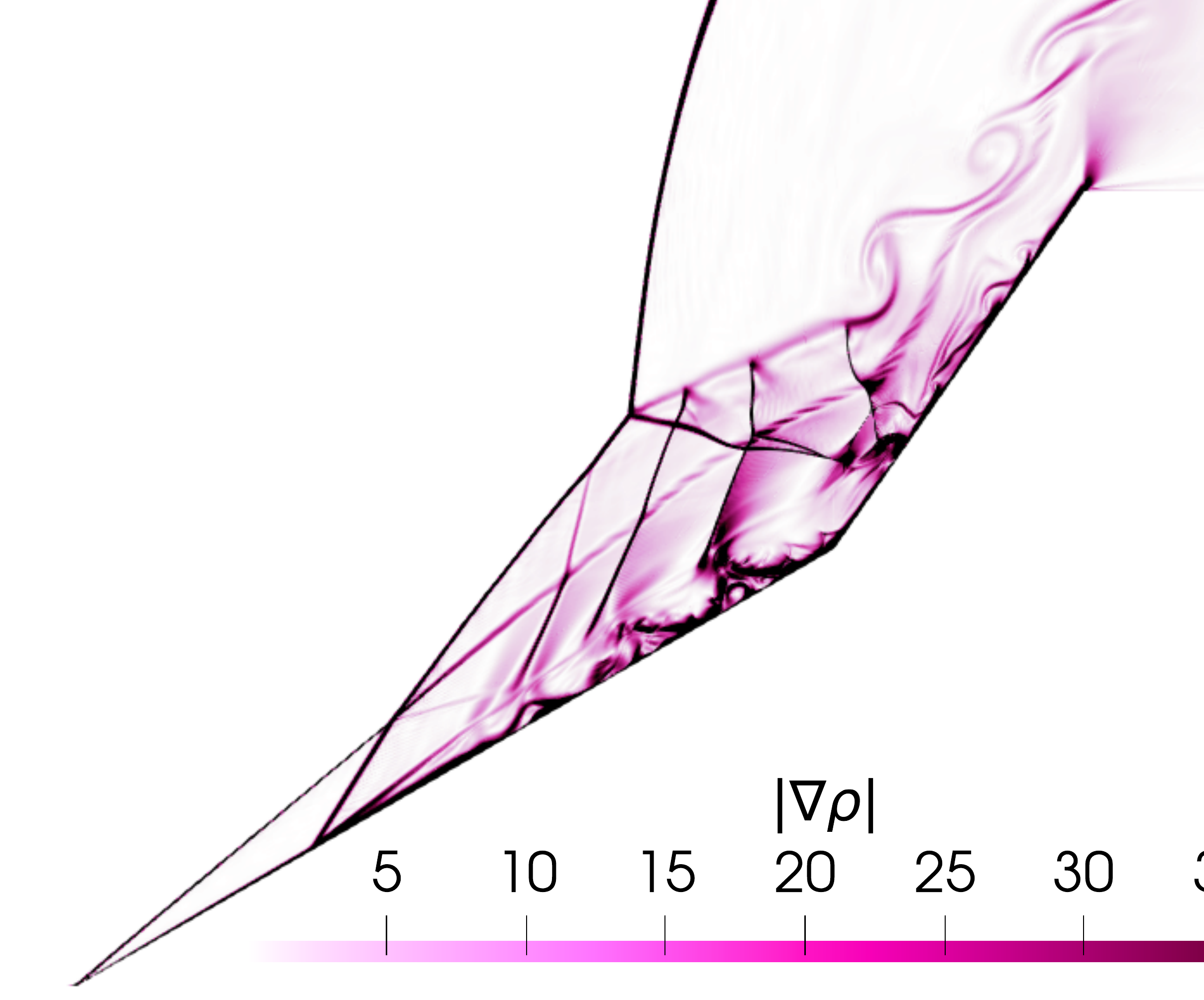}  
		\caption{}
		\label{fig:30_55_2_Finer_260us}
	\end{subfigure}
	\caption{A comparison of shock pattern resolution with four different grids with $\theta_2=55^\circ$ and $L_1/L_2=2$ at t = 260$\mu s$. (\subref{fig:30_55_2_swantek_260us}) Experiment (Reproduced from Swantek\cite{swantek2012heat} with permission from the author) (\subref{fig:30_55_2_Coarse_260us}) Coarse grid (\subref{fig:30_55_2_Medium_260us}) Medium grid (\subref{fig:30_55_2_Fine_260us}) Fine grid (\subref{fig:30_55_2_Finer_260us}) Finer grid. Visualization using schlieren images ($|\nabla \rho|$ in computation) }
	\label{fig:30_55_2_260us}
\end{figure}
It should be noted that the reattachment point is the same as the shock impingement location because the separation region is not yet fully developed, and the growth of the separation region is under the influence of adverse pressure gradient generated due to the transmitted shock (TS). But the further resolution of the grid from Medium to Fine where cell sizes are reduced by a factor of 5, the shifts in SP and RP are marginal. Further halving the cell size in finer grid does not provide any different results from the fine grid within engineering accuracy. The most important observation is the sharp resolution of the peak pressure on the aft-wedge and the sharpness of pressure peaks in the separated region. The pressure peak increases because of the increased strength of the transmitted shock resolved using refined grids. The improvement is shown due to grid resolution in the pattern of the wall pressure in fig. \ref{fig:pressure260usStudy1} and schlieren visualization in fig. \ref{fig:30_55_2_260us} complement each other.

\begin{figure}[h!]
	\centering
	\begin{subfigure}[t]{\linewidth}
		\centering
		\includegraphics[width=0.7\linewidth]{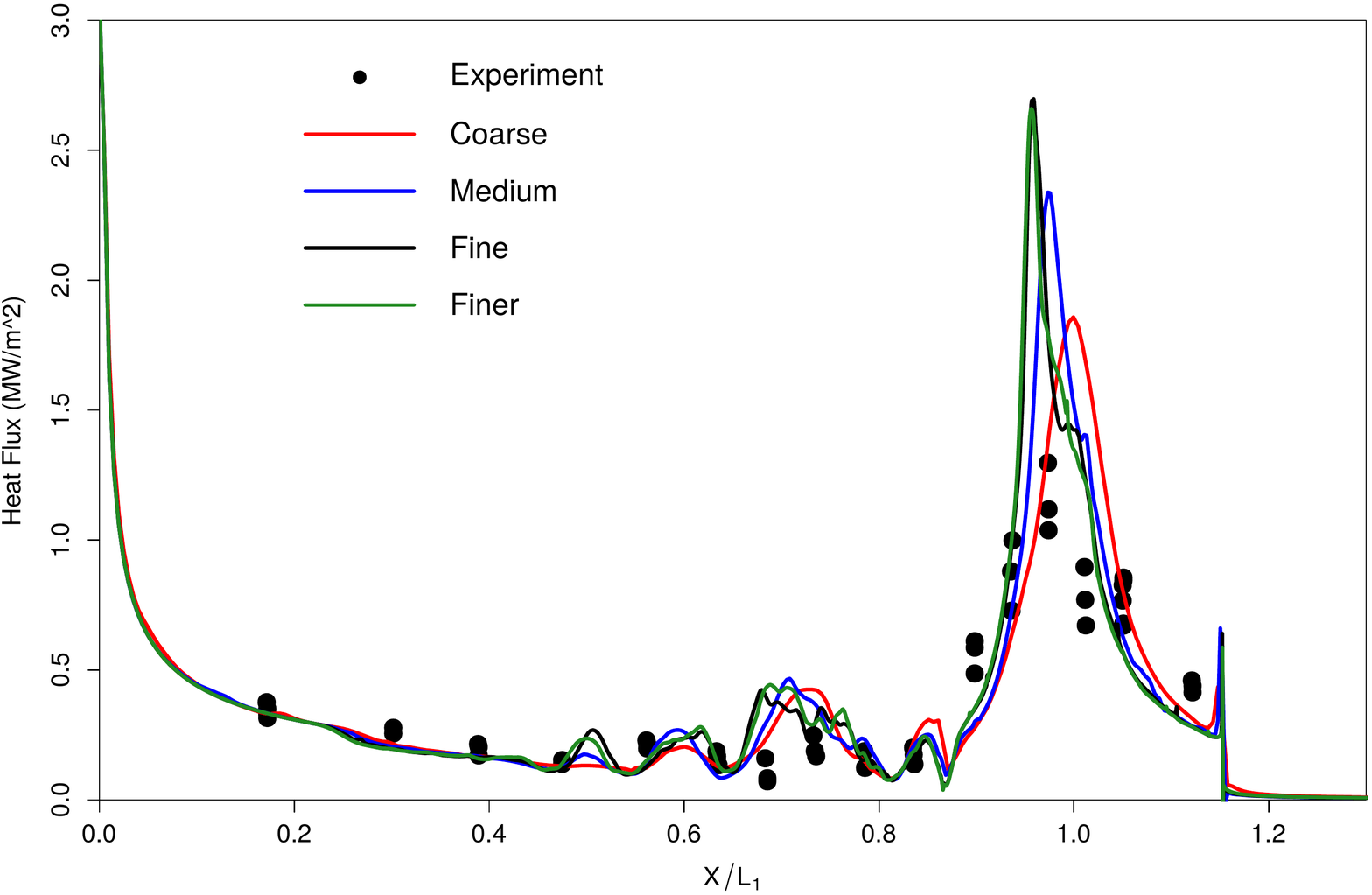}  
		\caption{}
		\label{fig:qMeanExpStudy1}
	\end{subfigure}
	\begin{subfigure}[t]{\linewidth}
		\centering
		\includegraphics[width=0.7\linewidth]{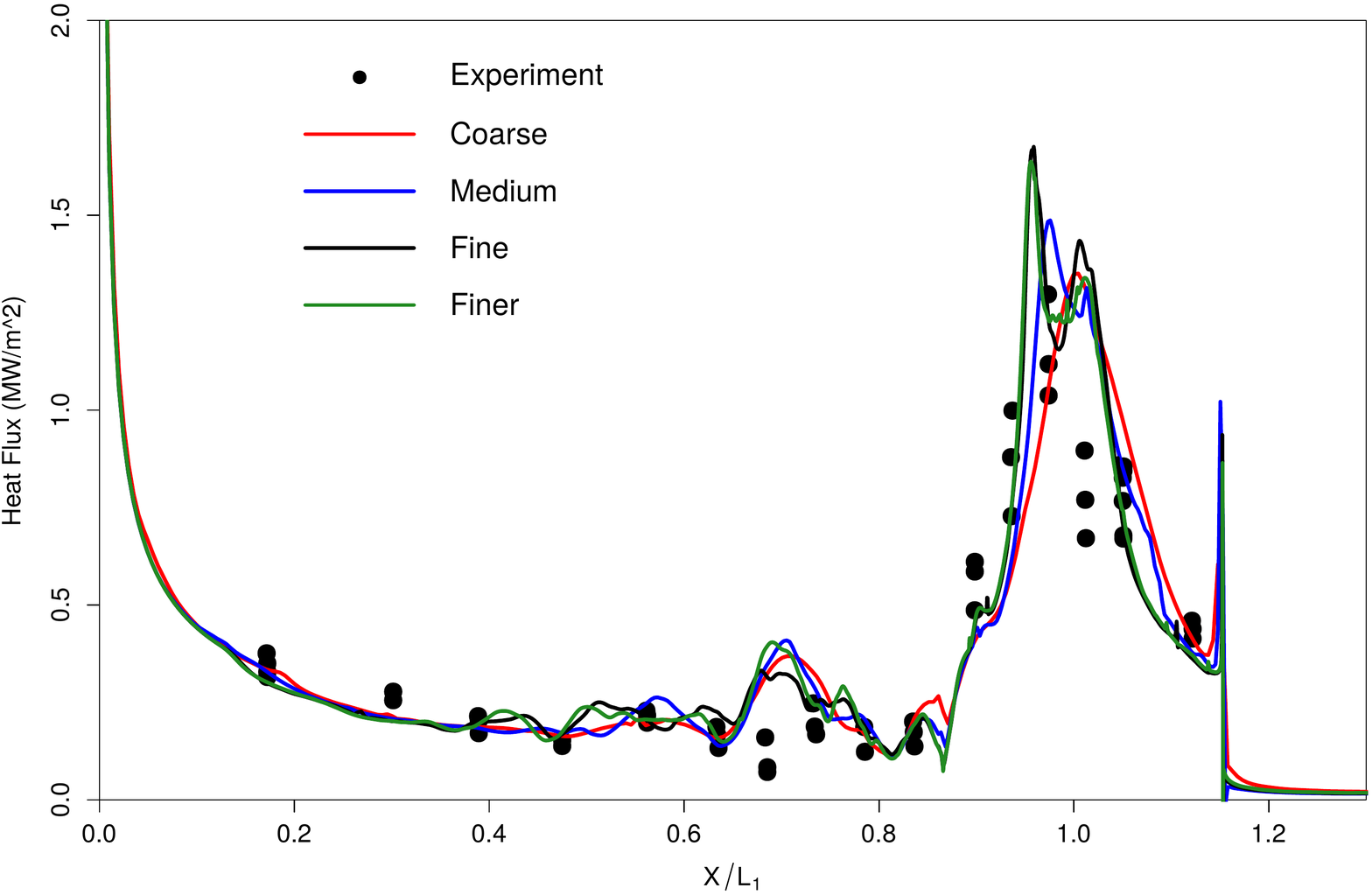}  
		\caption{}
		\label{fig:qMeanExp400usAvgStudy1}
	\end{subfigure}
	\caption{Mean wall heat flux with $\theta_2=55^\circ$ and $L_1/L_2=2$. (\subref{fig:qMeanExpStudy1}) Average between t = 150 $\mu s$ and 310 $\mu s$ (\subref{fig:qMeanExp400usAvgStudy1}) Average between t = 0 and 400 $\mu s$. Experimental data is from Swantek (Reproduced from Swantek\cite{swantek2012heat} with permission from the author)}.
	\label{fig:qMeanStudy1}
\end{figure}

Figure \ref{fig:qMeanStudy1} shows the comparison of average wall heat flux with the experimental results of Swantek\cite{swantek2012heat}. There is a discrepancy in the published literature regarding the time window of comparison between computed and experimental wall heat flux. In the experiments\cite{swantek2012heat}, average heat flux data is acquired after the flow has completely established. But the published computational research shows that the flow is not fully developed in the experimental duration of the measurement.
Following the averaging time window used in ref.\cite{durna2016shock,durna2019time}, figure \ref{fig:qMeanExpStudy1} shows the comparison of wall heat flux averaged between t = 150 $\mu s$ and 310 $\mu s$ for the four grids used in the simulation. All four grids show very similar results and are in good agreement with the experimental data from ref.\cite{swantek2012heat} except near the shock impingement point. With increased grid resolution, peak wall heat flux increases along with a movement upstream. The discrepancy between computation and experiment is attributed to the ambiguity in the time-averaging window used in the experiment. It is not surprising to see that a coarse grid predicts better peak wall heat flux compared to more resolved grids, but with the coarse grid, the separation region is also underpredicted with the incorrect resolution of shock waves and vortices. A similar observation was also made in ref.\cite{komives2014numerical} when using first-order numerical schemes. A time-averaging window of 400 $\mu s$ from the starting of the simulation, shown in fig. \ref{fig:qMeanExp400usAvgStudy1}, gives a very close match between the computed results and the experiment. In fig. \ref{fig:qMeanExp400usAvgStudy1}, near the shock impingement location on the aft-wedge, all the features with correct magnitude of the experimental wall heat flux data are closely replicated by the fine grid computation, and a smoothening of features is seen on a coarser grid. This behavior can be easily understood using the flow structure comparison shown in fig. \ref{fig:30_55_2_260us}.

Until now, the comparison is made in a small time window ($\approx$ 310 $\mu s$) in accordance with the duration of the experiment; but, the computational studies in the literature reveal that the complete flow development takes much longer than the experimental duration and the simulations need to be run for a longer time. It has been seen that the small separation region formed in fig. \ref{fig:30_55_2_260us} is not fully developed. Though all three grids give small variation in overall flow at the early stages of flow development, after the flow is fully developed, a large separation region will be formed. In fig. \ref{fig:30_55_2_8ms}, a comparison between the four grids is shown at t = 8ms when the flow is fully developed. It also shows the time-dependent variation of wall pressure distribution along the surface for the three grids. In figures \ref{fig:30_55_2_Coarse_pressure}, \ref{fig:30_55_2_Medium_pressure} and \ref{fig:30_55_2_Fine_pressure} pressure contours are shown in x-t dimensions. The X-axis represents the horizontal location on the wedge surface, whereas the y-axis represents the physical time of the flow. X/L = 0 is the location of the leading edge. Due to a large variation in pressure at the separation point, reattachment point, or shock impingement points, pressure iso-contours can be used in tracing the movement of such flow structures in space and time simultaneously. 

\begin{figure}[H]
	\centering
	\includegraphics[width=\linewidth, trim={0 0 0 12cm}, clip]{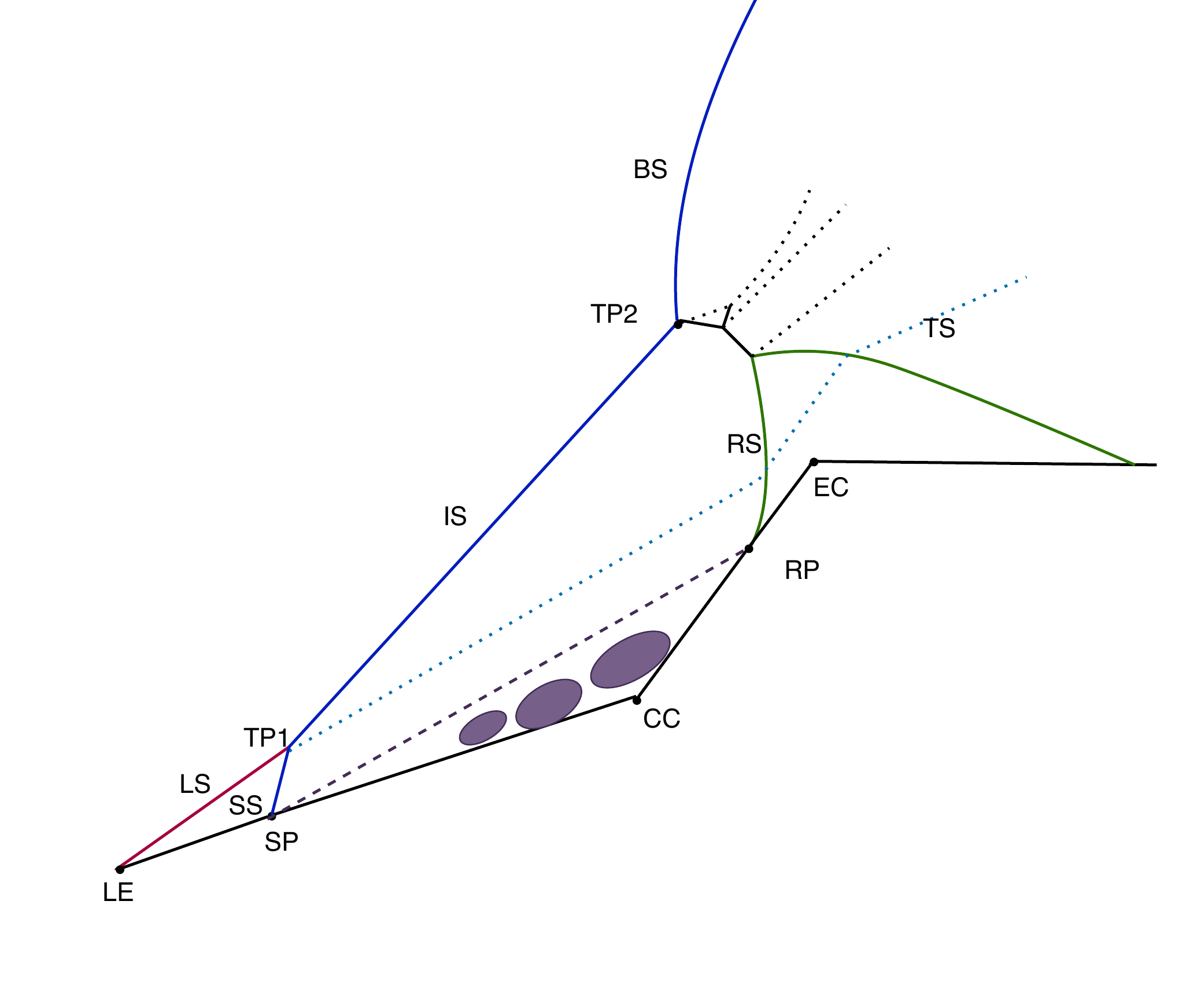}  
	\caption{A schematic of shock-shock interaction and shock reflection pattern with a large separation region. Nomenclature: LE (Leading edge), CC (Compression corner), EC (Expansion corner), SP (Separation point), RP (Reattachment point), TP1 (First tripple point), TP2 (Second tripple point), LS(Leading edge shock), SS (Separation shock), IS (Intermediate shock), BS (Bow shock), RS (Reattachment shock), TS (Transmitted shock). Dashed lines represent the shear layer.}
	\label{fig:sch3ms}
\end{figure}

\begin{figure}[h!]
	\centering
	\begin{subfigure}[t]{\linewidth}
		\centering
		\includegraphics[width=\linewidth, trim={1cm 20cm 0 0}, clip]{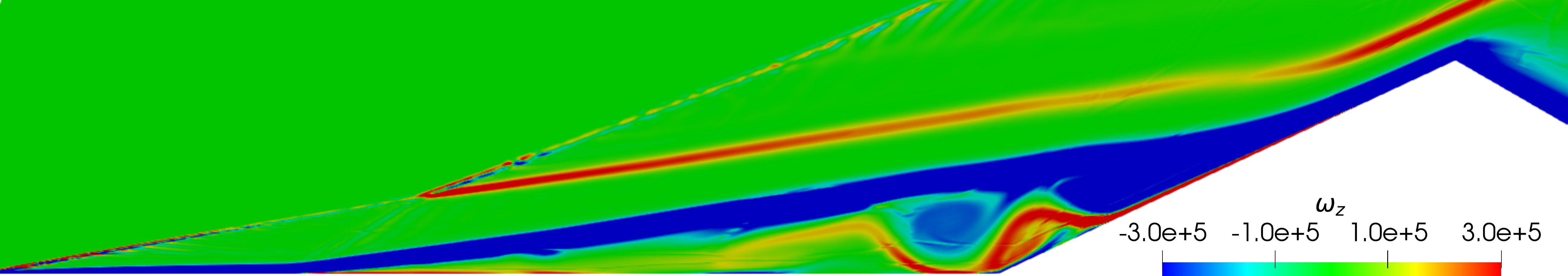}  
		\caption{}
		\label{fig:Medium_vorticity}
	\end{subfigure}
	\begin{subfigure}[t]{\linewidth}
		\centering
		\includegraphics[width=\linewidth, trim={1cm 20cm 0 0}, clip]{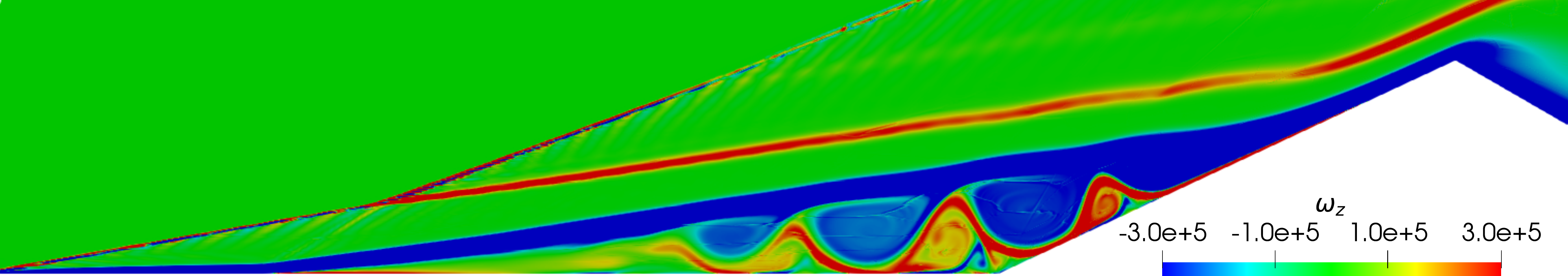}  
		\caption{}
		\label{fig:Fine_vorticity}
	\end{subfigure}
	\begin{subfigure}[t]{\linewidth}
		\centering
		\includegraphics[width=\linewidth, trim={1cm 20cm 0 0}, clip]{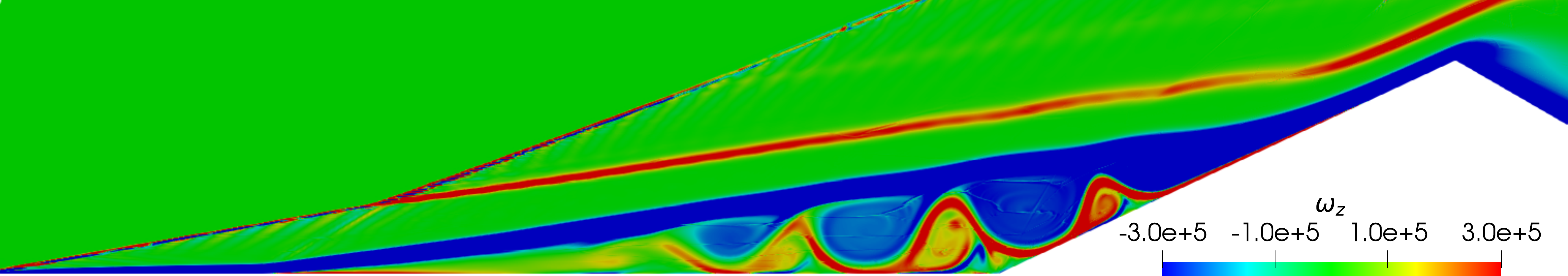}  
		\caption{}
		\label{fig:Finer_vorticity}
	\end{subfigure}
	\caption{Comparison of span-wise vorticity in the corner separated flow region with $\theta_2 = 55^\circ$ and $L_1/L_2=2$ due to grid resolution. (\subref{fig:Medium_vorticity}) Medium grid (\subref{fig:Fine_vorticity}) Fine grid (\subref{fig:Finer_vorticity}) Finer grid}
	\label{fig:30_55_2_7p04ms_vorticity}
\end{figure}

In fig. \ref{fig:30_55_2_Coarse_8ms} and \ref{fig:30_55_2_Medium_8ms}, the resolution of shock strengths and vorticity has improved with grid resolution so that the overall size of the separation region is larger on the medium grid compared to the coarse grid. This has also pushed the shock-shock interaction region upwards, and the transmitted shock no longer impinges near the expansion corner for the medium grid. A schematic for this shock-shock interaction pattern is shown in fig. \ref{fig:sch3ms}. This defers from the shock pattern on the aft-wedge shown in fig. \ref{fig:sch260us} in two ways. First, the transmitted shock (TS) no longer impinges on the aft-wedge and has moved downstream of the expansion corner. Second, the tripple point (TP2) has moved farther away from the wedge so that a type V shock-shock interaction pattern is formed.
\begin{figure}[h!]
	\centering
	\begin{subfigure}[t]{0.545\linewidth}
		\centering
		\includegraphics[width=\linewidth, trim={0 0 10cm 0}, clip]{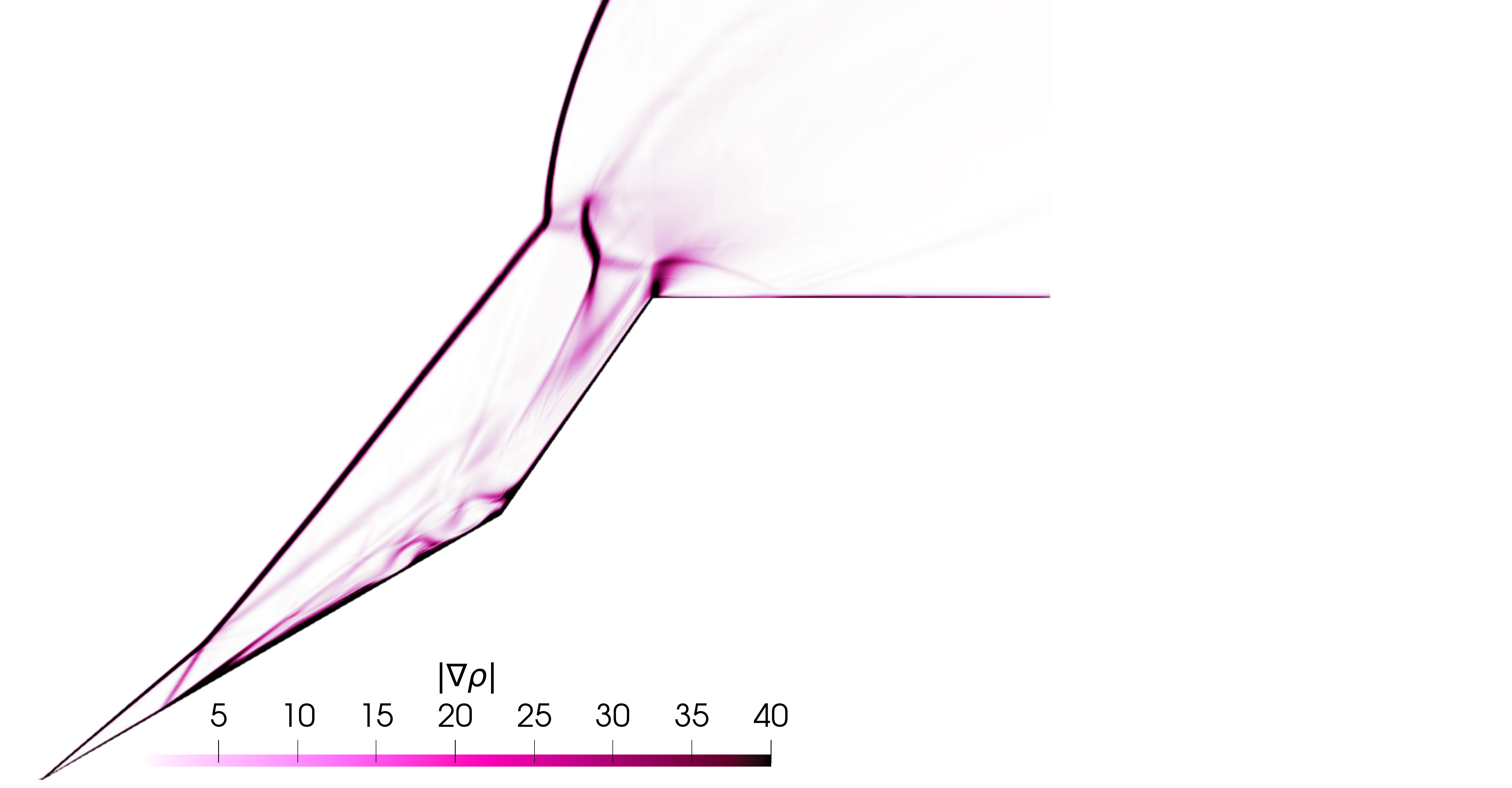}  
		\caption{}
		\label{fig:30_55_2_Coarse_8ms}
	\end{subfigure}
	\begin{subfigure}[t]{0.445\linewidth}
		\centering
		\includegraphics[width=\linewidth, trim={0 0 20cm 0}, clip]{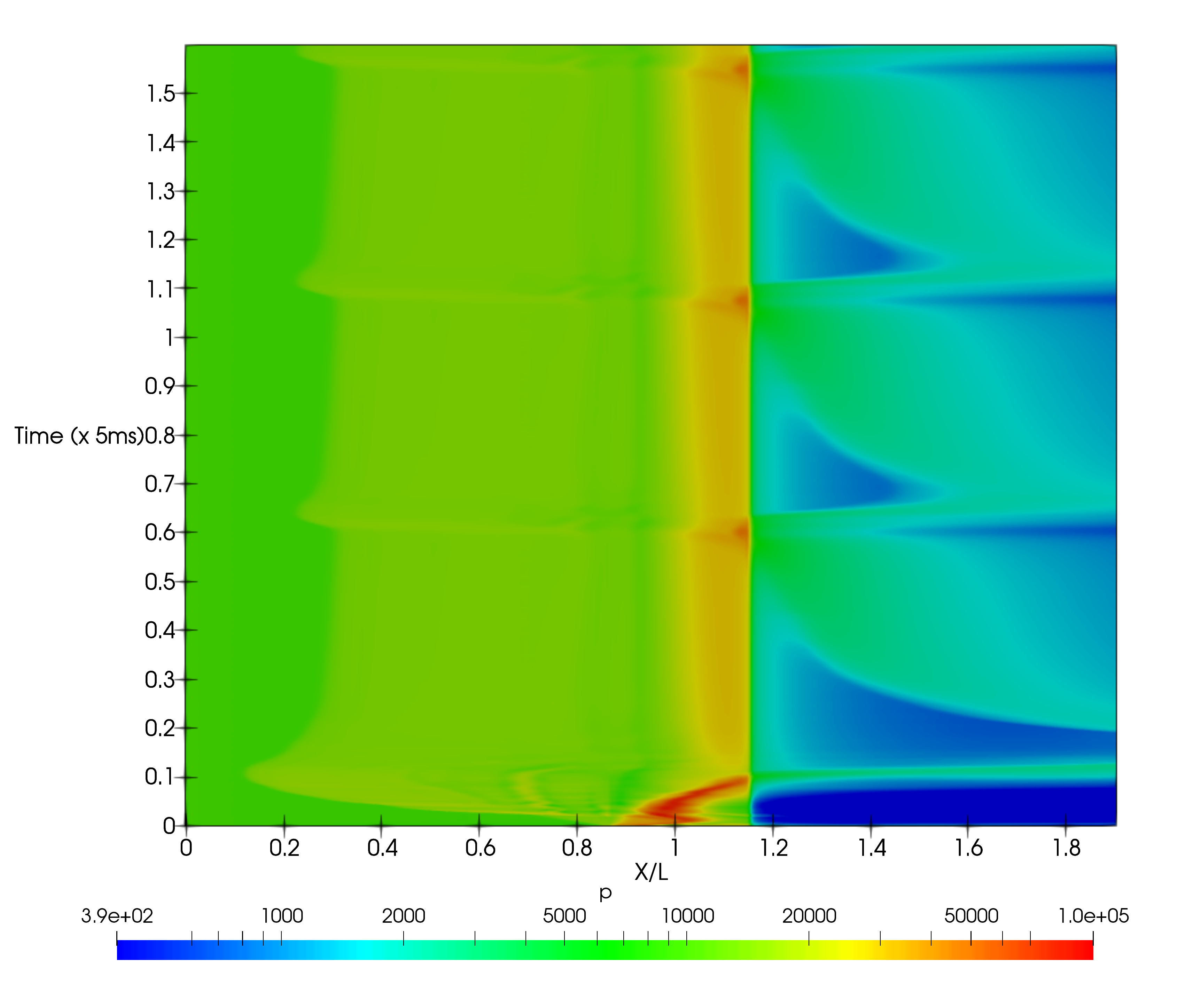}  
		\caption{}
		\label{fig:30_55_2_Coarse_pressure}
	\end{subfigure}
	\begin{subfigure}[t]{0.545\linewidth}
		\centering
		\includegraphics[width=\linewidth, trim={0 0 10cm 0}, clip]{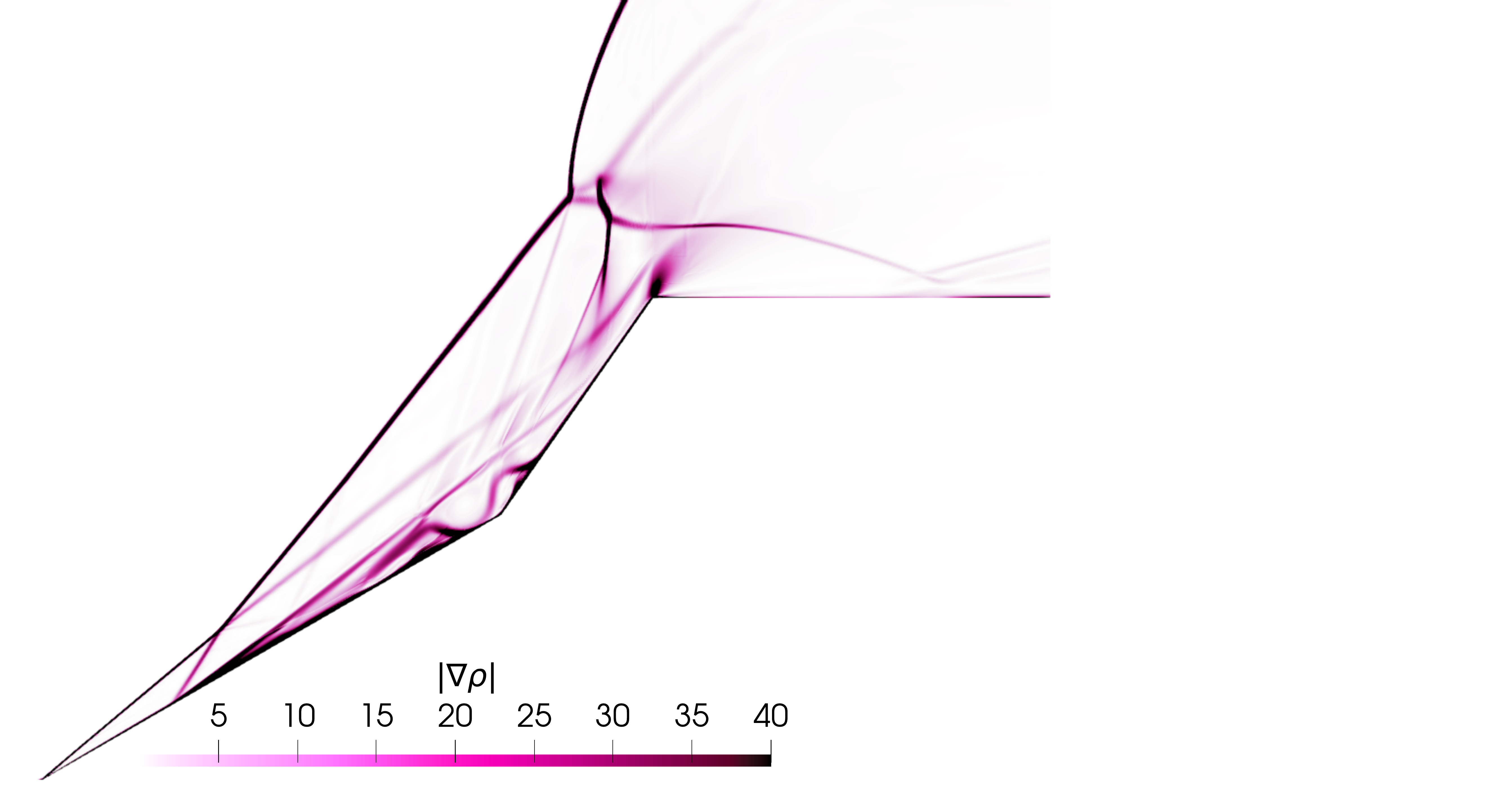}  
		\caption{}
		\label{fig:30_55_2_Medium_8ms}
	\end{subfigure}
	\begin{subfigure}[t]{0.445\linewidth}
		\centering
		\includegraphics[width=\linewidth, trim={0 0 20cm 0}, clip]{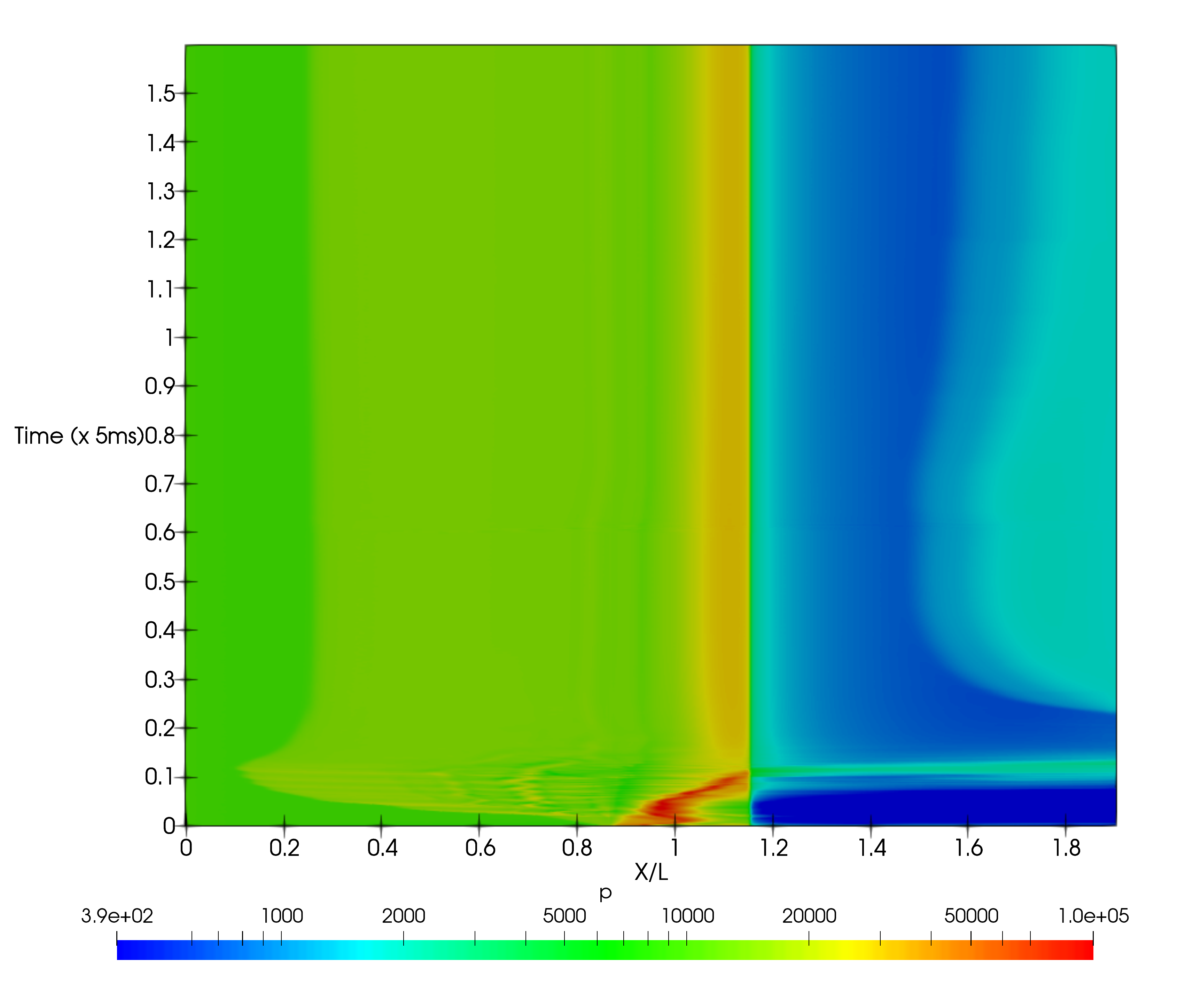}  
		\caption{}
		\label{fig:30_55_2_Medium_pressure}
	\end{subfigure}
\end{figure}
\begin{figure}[h!]\ContinuedFloat
	\begin{subfigure}[t]{0.545\linewidth}
		\centering
		\includegraphics[width=\linewidth, trim={0 0 10cm 0}, clip]{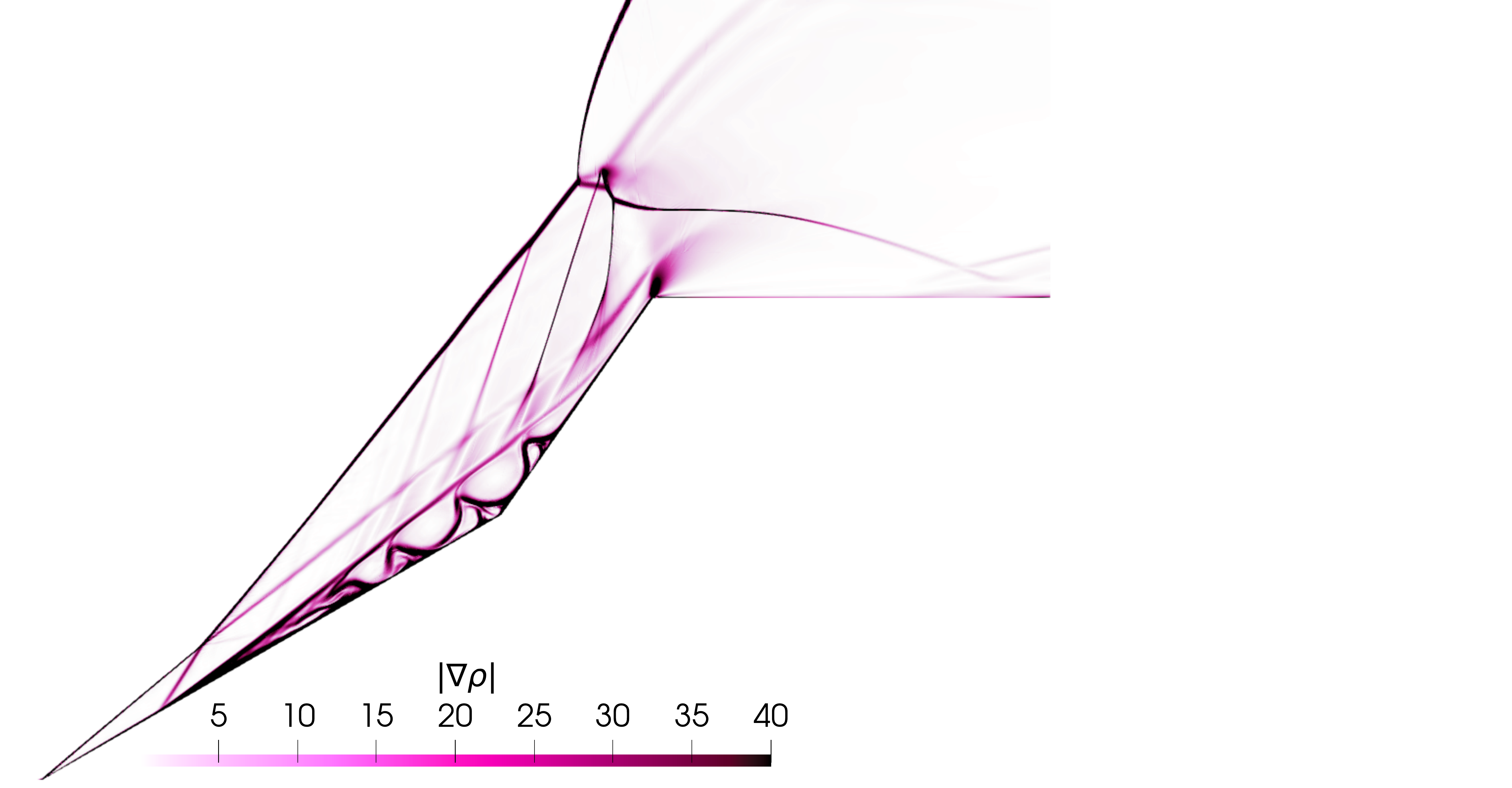}  
		\caption{}
		\label{fig:30_55_2_Fine_8ms}
	\end{subfigure}
	\begin{subfigure}[t]{0.445\linewidth}
		\centering
		\includegraphics[width=\linewidth, trim={0 0 20cm 0}, clip]{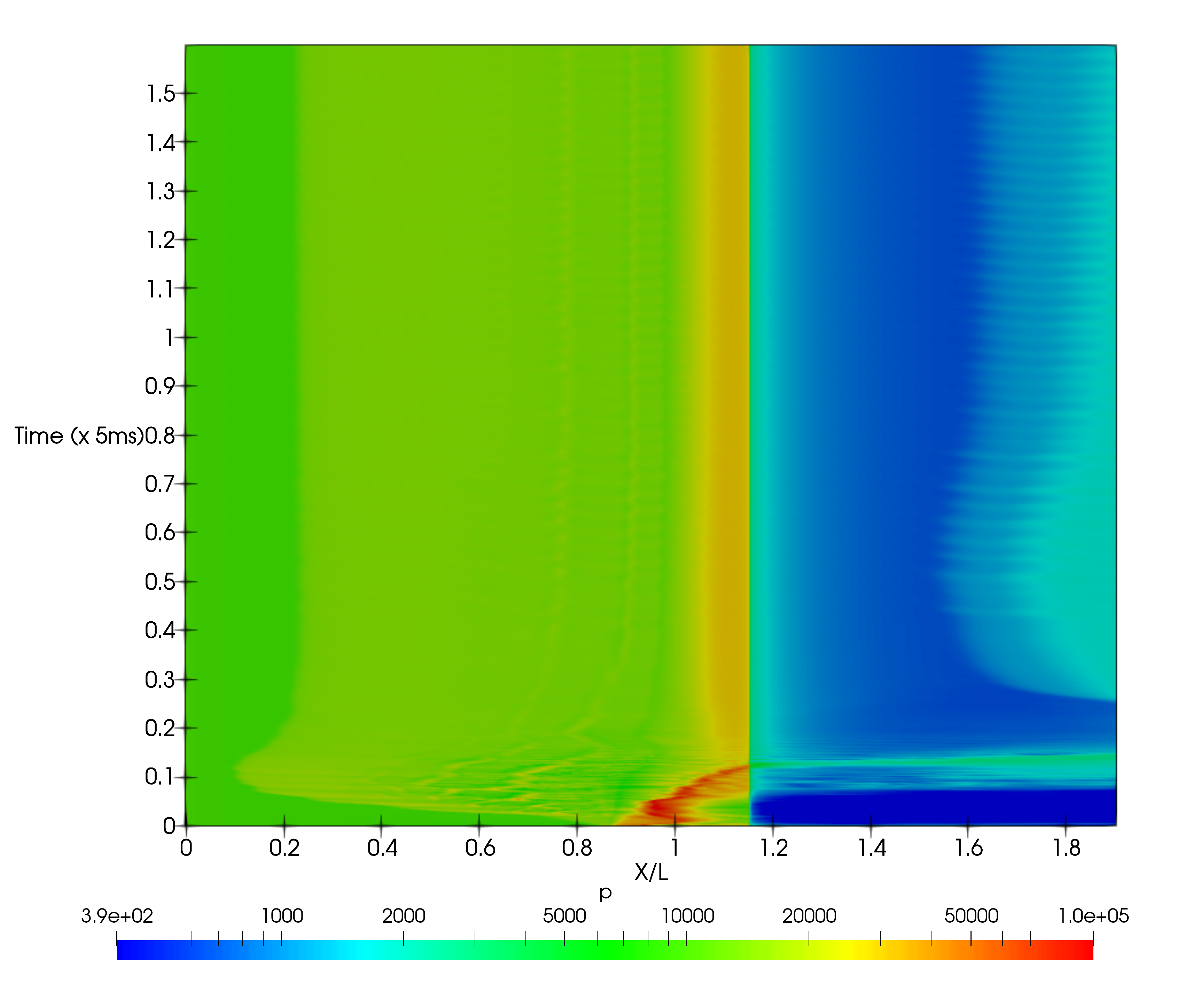}  
		\caption{}
		\label{fig:30_55_2_Fine_pressure}
	\end{subfigure}
	\begin{subfigure}[t]{0.545\linewidth}
		\centering
		\includegraphics[width=\linewidth, trim={0 0 10cm 0}, clip]{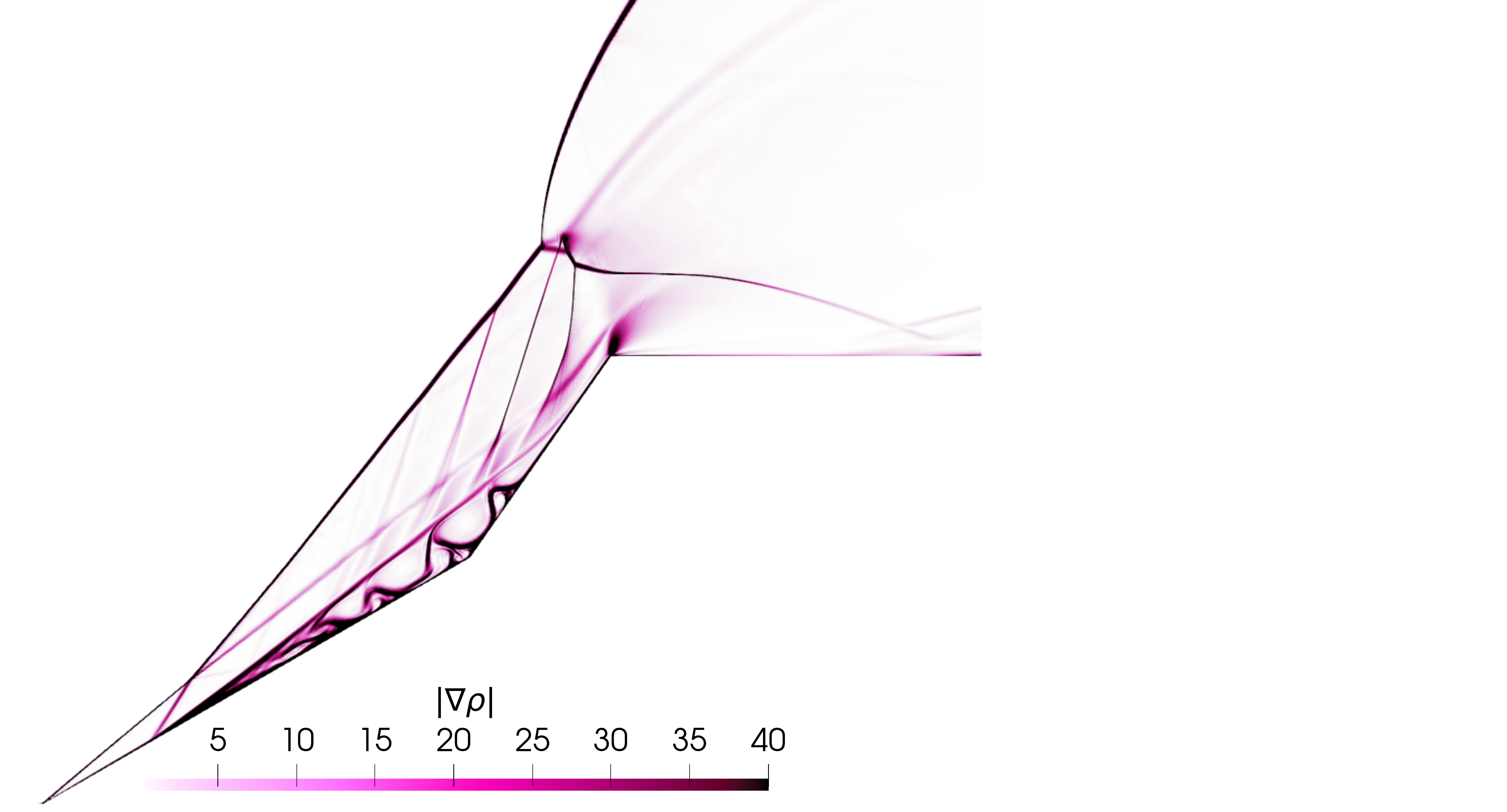}  
		\caption{}
		\label{fig:30_55_2_Finer_8ms}
	\end{subfigure}
	\begin{subfigure}[t]{0.445\linewidth}
		\centering
		\includegraphics[width=\linewidth, trim={0 0 20cm 0}, clip]{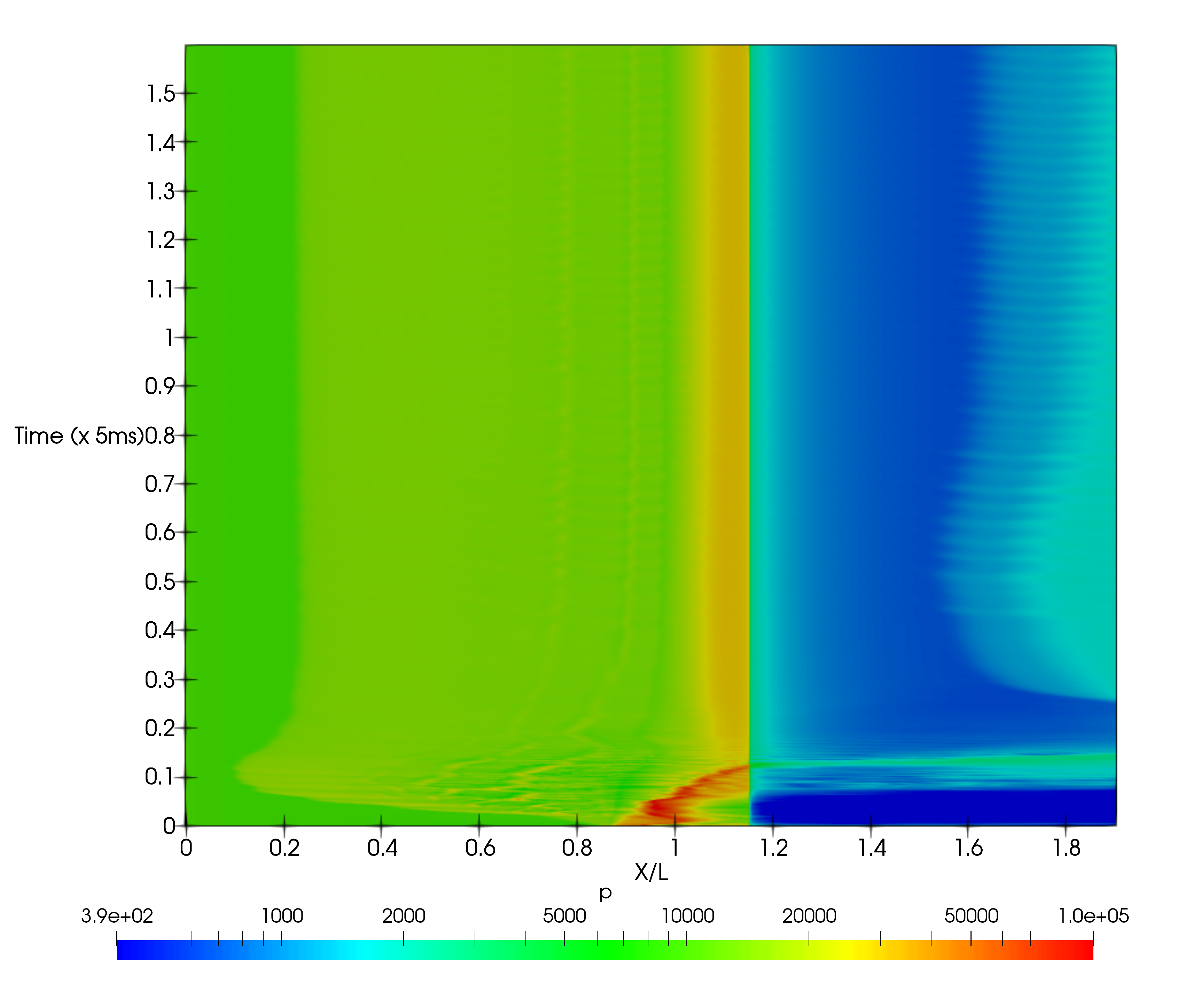}  
		\caption{}
		\label{fig:30_55_2_Finer_pressure}
	\end{subfigure}
	\caption{Numerical schlieren visualization of shock waves and separation region and spatio-temporal variation of wall pressure at $L_1/L_2=2$ and $\theta_2 = 55^\circ$ with different grid resolutions. (\subref{fig:30_55_2_Coarse_8ms}, \subref{fig:30_55_2_Coarse_pressure}) Coarse grid  (\subref{fig:30_55_2_Medium_8ms}, \subref{fig:30_55_2_Medium_pressure}) Medium grid ( \subref{fig:30_55_2_Fine_8ms}, \subref{fig:30_55_2_Fine_pressure}) Fine grid (\subref{fig:30_55_2_Finer_8ms}, \subref{fig:30_55_2_Finer_pressure}) Finer grid }
	\label{fig:30_55_2_8ms}
\end{figure}
This has a significant impact on the overall unsteady flow physics in this case. In fig. \ref{fig:30_55_2_Coarse_pressure}, it can be seen that the pressure distribution on the wedge periodically changes with time due to a sustained oscillation of the shock structure over the second wedge. This has also been studied in detail in the previous literature\cite{durna2016shock,durna2019time}. But in \ref{fig:30_55_2_Medium_pressure}, flow is seen to reach a steady state because of improvement in the prediction of separation region size due to better grid resolution. Further resolutions of the grid in fine and finer grid cases have revealed a series of primary, secondary, and tertiary vortices confined within the separation region, which shows unsteady motion and produces small time-dependent oscillation in the pressure distribution as shown in fig. \ref{fig:30_55_2_Fine_pressure}. More detailed vorticity pattern comparison between medium, fine and finer grid is shown in fig. \ref{fig:30_55_2_7p04ms_vorticity}. In fig. \ref{fig:Medium_vorticity}, only one primary and one secondary vortices are generated at the compression corner, which reaches a steady state after the separation region is fully developed. A similar vorticity pattern was also seen in 3D simulations of ref.\cite{komives2014numerical} But on further resolving the grid in fig. \ref{fig:Fine_vorticity} and \ref{fig:Finer_vorticity}, more primary and secondary vortices are seen in the separation region along with a pair of oscillating tertiary vortices, which makes the complete flow unsteady.

Hence, it is shown that the fine grid is best suited for long-duration flow simulation in this case. The medium grid produces results very close to Fine grid, but the high-frequency oscillation of the shock-system generated due to vortical motion inside the separation region is lost. The coarse grid cannot be used because it produces a result similar to other grids at the beginning of flow development, but in the long term, it produces a spurious low-frequency shock oscillation due to incorrect prediction of separation region and transmitted shock. And, the finer grid produces the same results as the fine grid without any significant improvement. This shows that the fine grid is the converged grid and is sufficient for correctly resolving such high speed separated flows for a very long time.

\subsection{Study 2}
\begin{figure}[h!]
	\centering
	\begin{subfigure}[t]{\linewidth}
		\centering
		\includegraphics[width=0.7\linewidth]{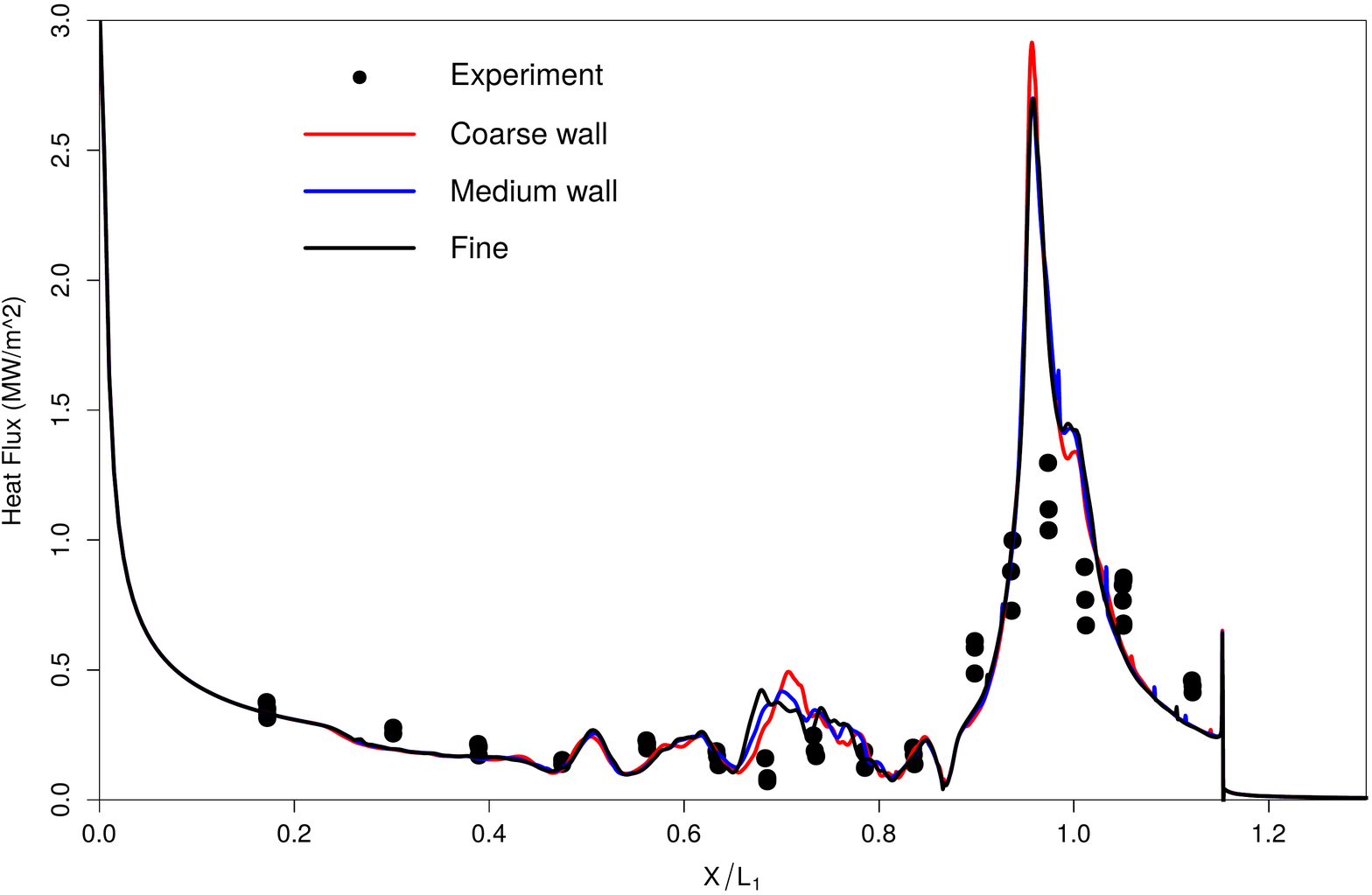}  
		\caption{}
		\label{fig:qMeanExpStudy2}
	\end{subfigure}
	\begin{subfigure}[t]{\linewidth}
		\centering
		\includegraphics[width=0.7\linewidth]{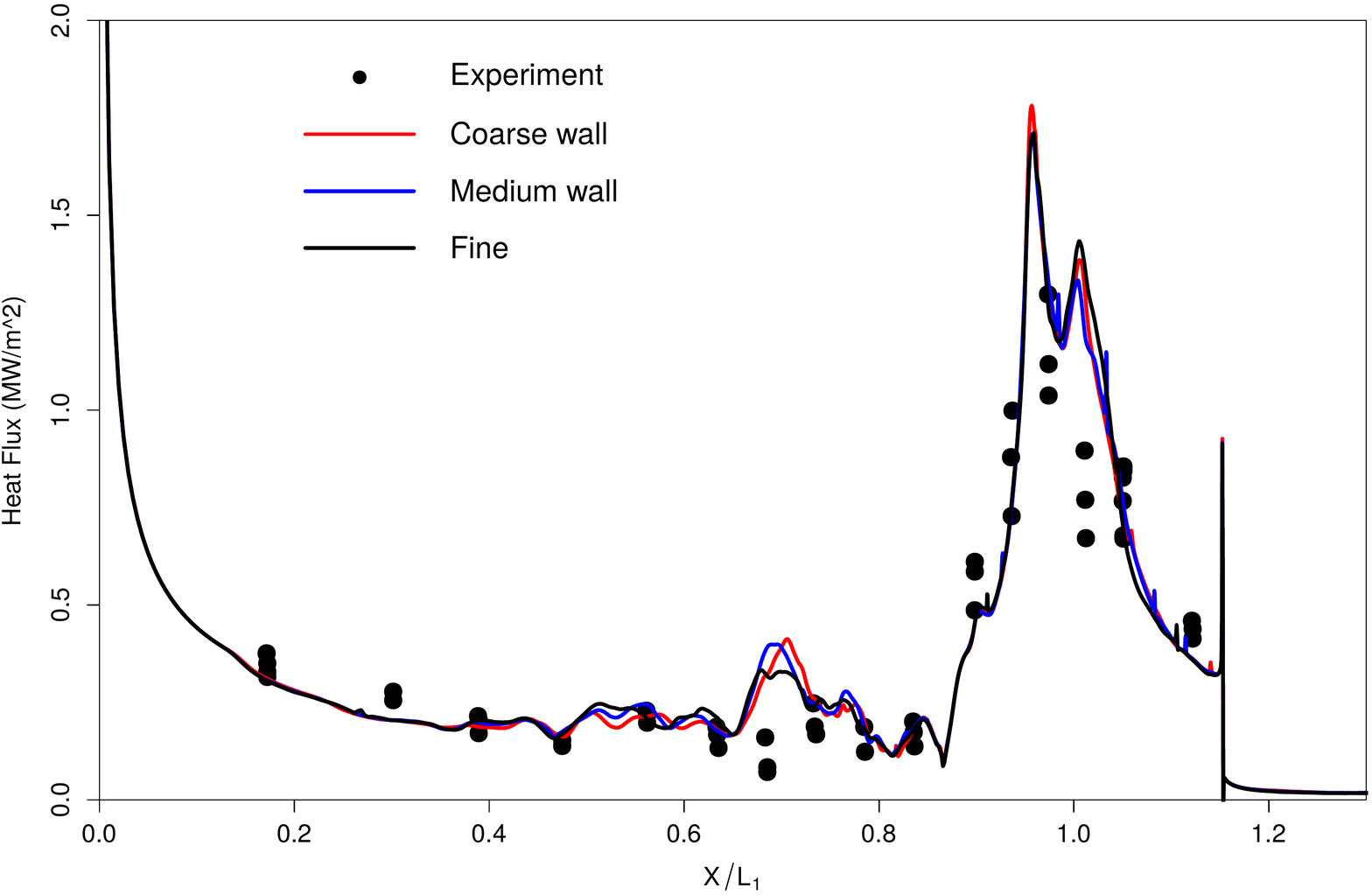}  
		\caption{}
		\label{fig:qMeanExp400usAvgStudy2}
	\end{subfigure}
	\caption{Mean wall heat flux with $\theta_2=55^\circ$ and $L_1/L_2=2$. (\subref{fig:qMeanExpStudy2}) Average between t = 150 $\mu s$ and 310 $\mu s$ (\subref{fig:qMeanExp400usAvgStudy2}) Average between t = 0 and 400 $\mu s$. Experimental data is from Swantek (Reproduced from Swantek\cite{swantek2012heat} with permission from the author)} 
	\label{fig:qMeanStudy2}
\end{figure}

\begin{figure}[h!]
	\centering
	\includegraphics[width=0.7\linewidth]{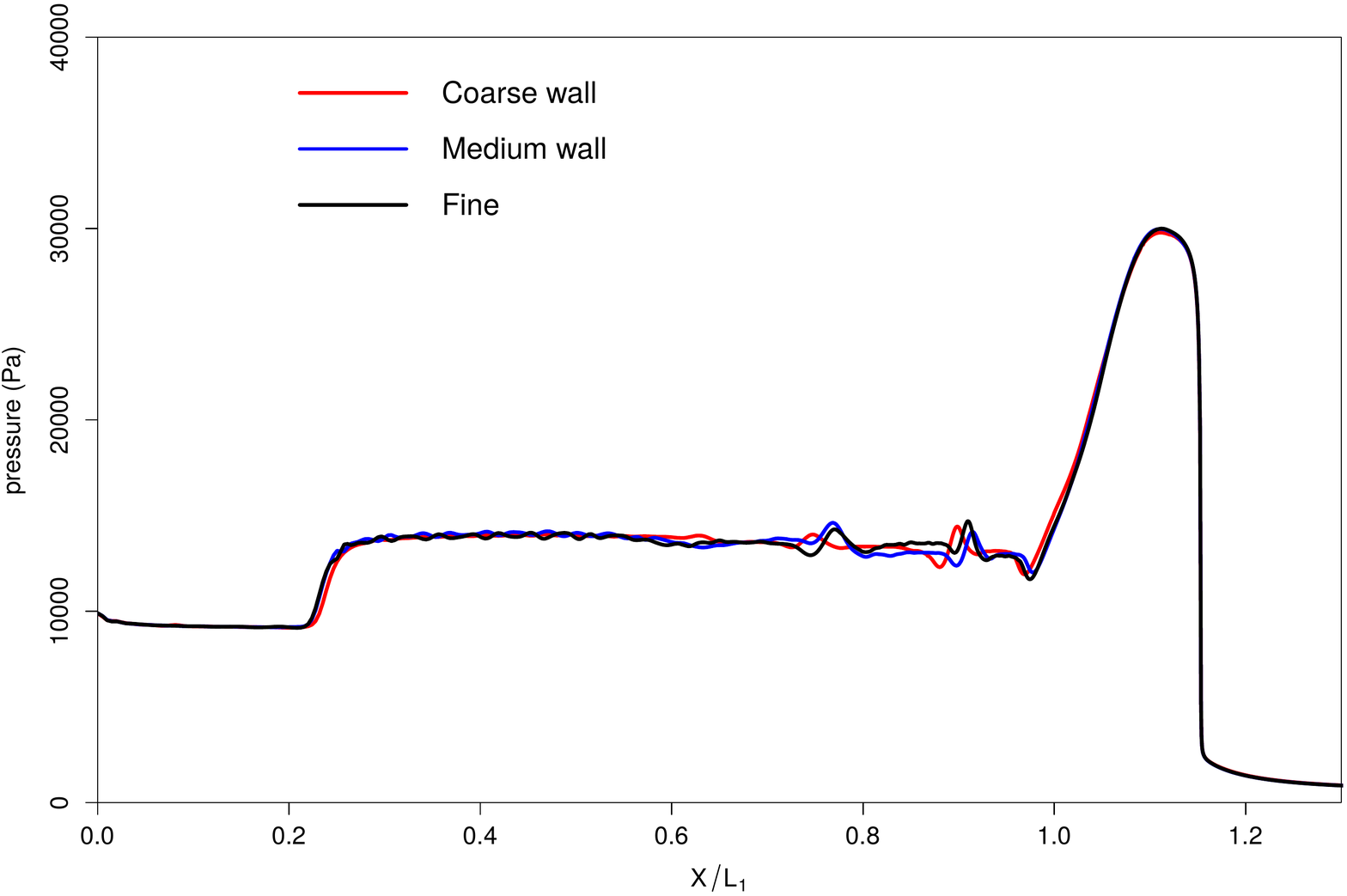}  
	\caption{Instantaneous wall pressure distribution with $\theta_2=55^\circ$ and $L_1/L_2=2$ at t = 3 ms}
	\label{fig:pressure3msStudy2}
\end{figure}

In the previous grid comparison study, the grid had been resolved, maintaining the wall and the boundary cells resolution the same. But wall resolution can also affect the separation point location and shock boundary layer interaction pattern. Hence in this subsection, two additional grids with first grid cell height at the wall = 2$\times10^{-4}\times L_1$ (Medium wall grid), and 4$\times10^{-4}\times L_1$ (Coarse wall grid) has been considered along with fine grid ($10^{-4}\times L_1$). Averaged wall heat flux between t = 150 $\mu s$ and 310 $\mu s$ is shown in fig. \ref{fig:qMeanExpStudy2} and between t = 0 and 400 $\mu s$ is fig. \ref{fig:qMeanExp400usAvgStudy2}. Both figures show that Medium and fine wall resolution produces the same result, and a slightly increased peak value of wall heat flux is observed for coarse wall grid. Also, instantaneous wall pressure distribution along the wedge is shown in fig.\ref{fig:pressure3msStudy2} for t = 3ms. A negligible variation in the pressure distribution is seen, and the grid independence study can be considered converged for the fine grid solutions. This should also be noted that the overall flow physics is much more sensitive to the grid resolution outside the boundary layer as compared to inside the boundary at the wall due to the presence of a large separation bubble.

\section{Results and Discussion\label{sec:result}}
In the previous section, we saw that with an insufficient grid, an underpredicted separation region could result in the wrong locations of shock structures. If this results in the incidence shock impinging on the aft-wedge, there could be a spurious unsteady and periodic shock-shock interaction, which was a topic of discussion in ref.\cite{durna2016shock,durna2019time}. However, a further grid resolved simulation on the fine grid has shown that the shock-shock interaction becomes quasi-steady with a high-frequency oscillation due to the tertiary vortices.  

\begin{figure}[h!]
	\centering
	\begin{subfigure}[t]{0.495\linewidth}
		\centering
		\includegraphics[width=\linewidth, trim={0 0 10cm 0}, clip]{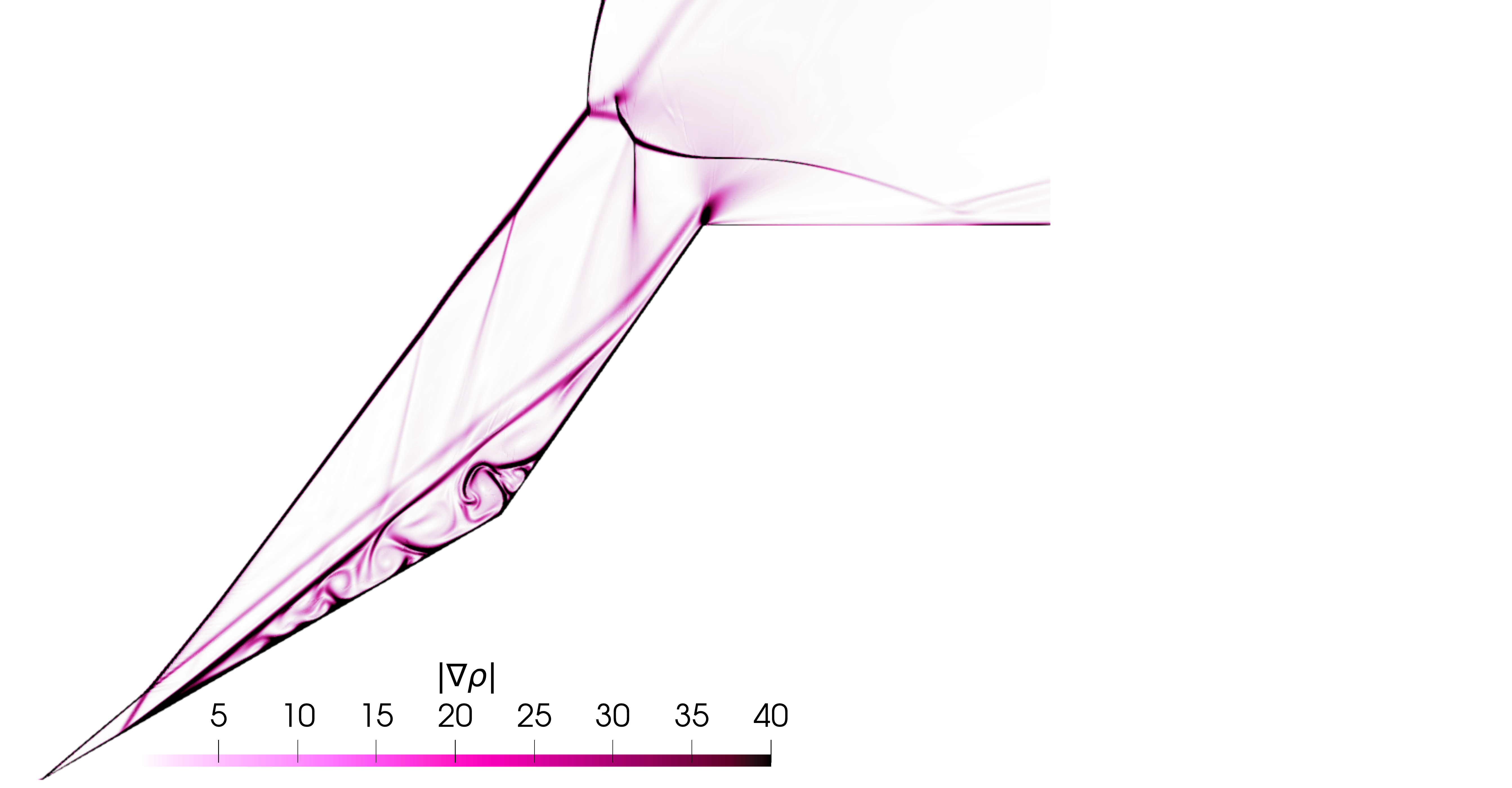}  
		\caption{}
		\label{fig:30_55_1p5_6p3ms}
	\end{subfigure}
	\begin{subfigure}[t]{0.495\linewidth}
		\centering
		\includegraphics[width=\linewidth, trim={0 0 10cm 0}, clip]{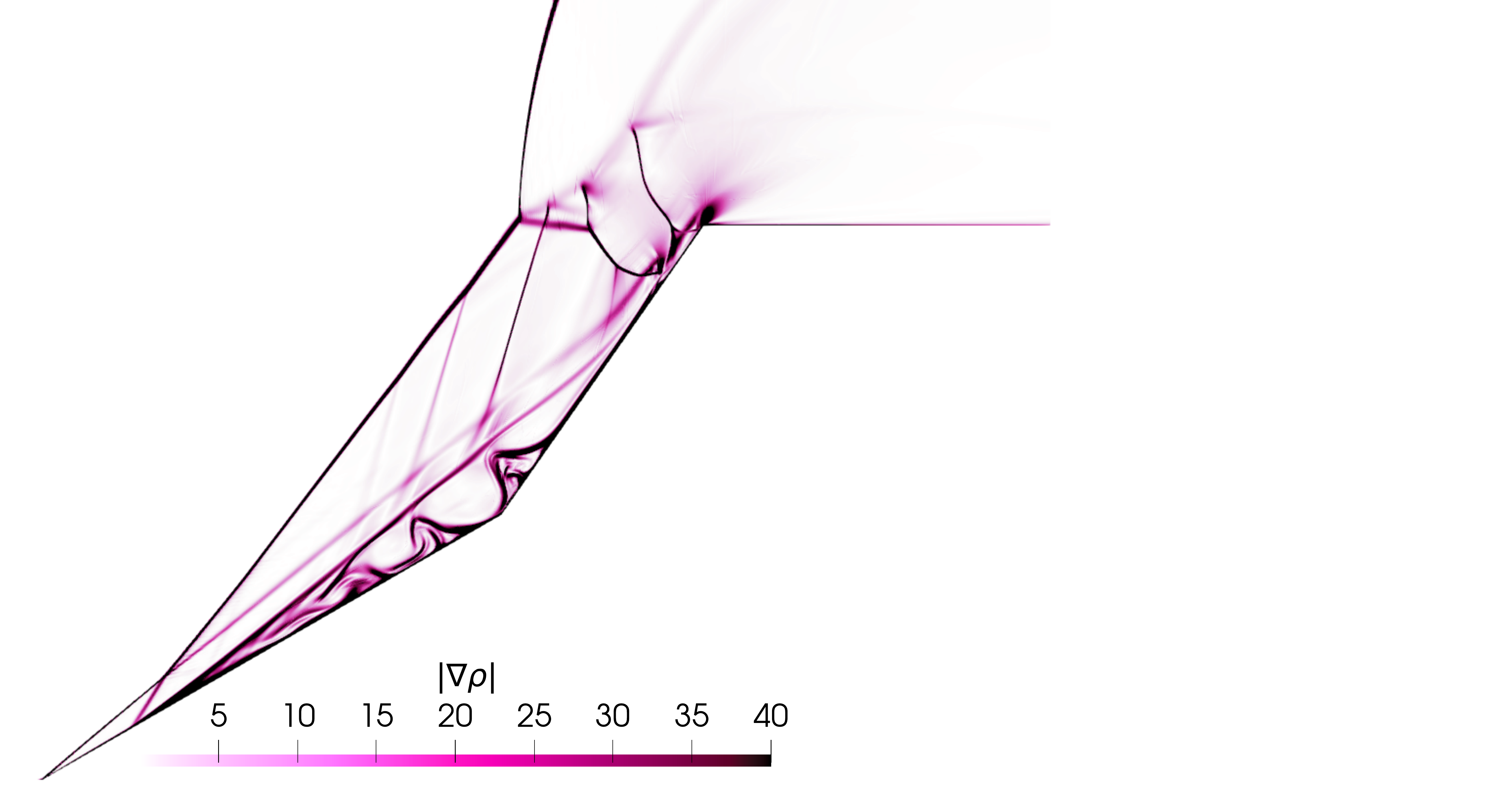}  
		\caption{}
		\label{fig:30_55_1p5_7ms}
	\end{subfigure}
	\begin{subfigure}[t]{0.495\linewidth}
		\centering
		\includegraphics[width=\linewidth, trim={0 0 10cm 0}, clip]{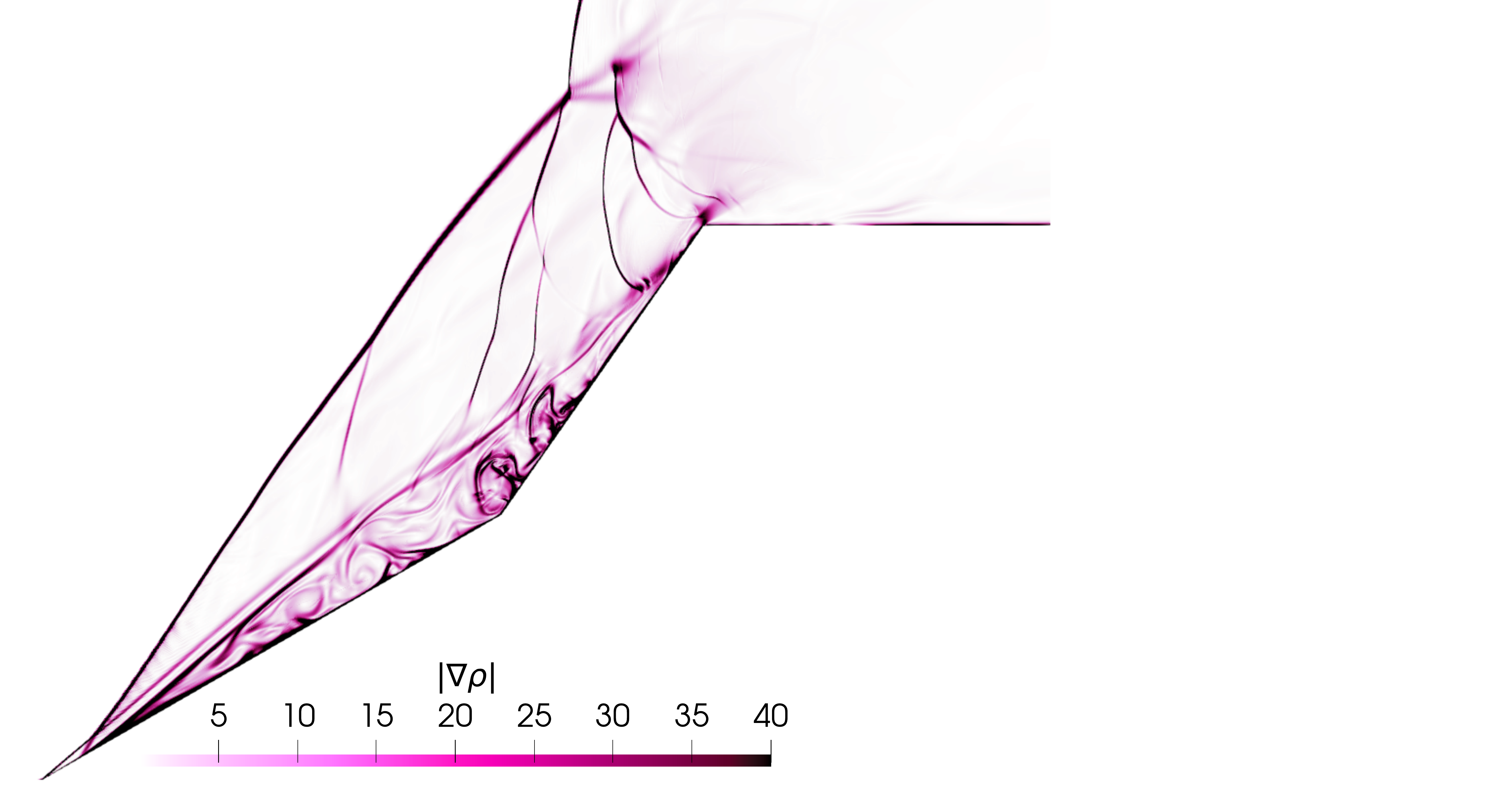}  
		\caption{}
		\label{fig:30_55_1p5_7p5ms}
	\end{subfigure}
	\begin{subfigure}[t]{0.495\linewidth}
		\centering
		\includegraphics[width=\linewidth, trim={0 0 10cm 0}, clip]{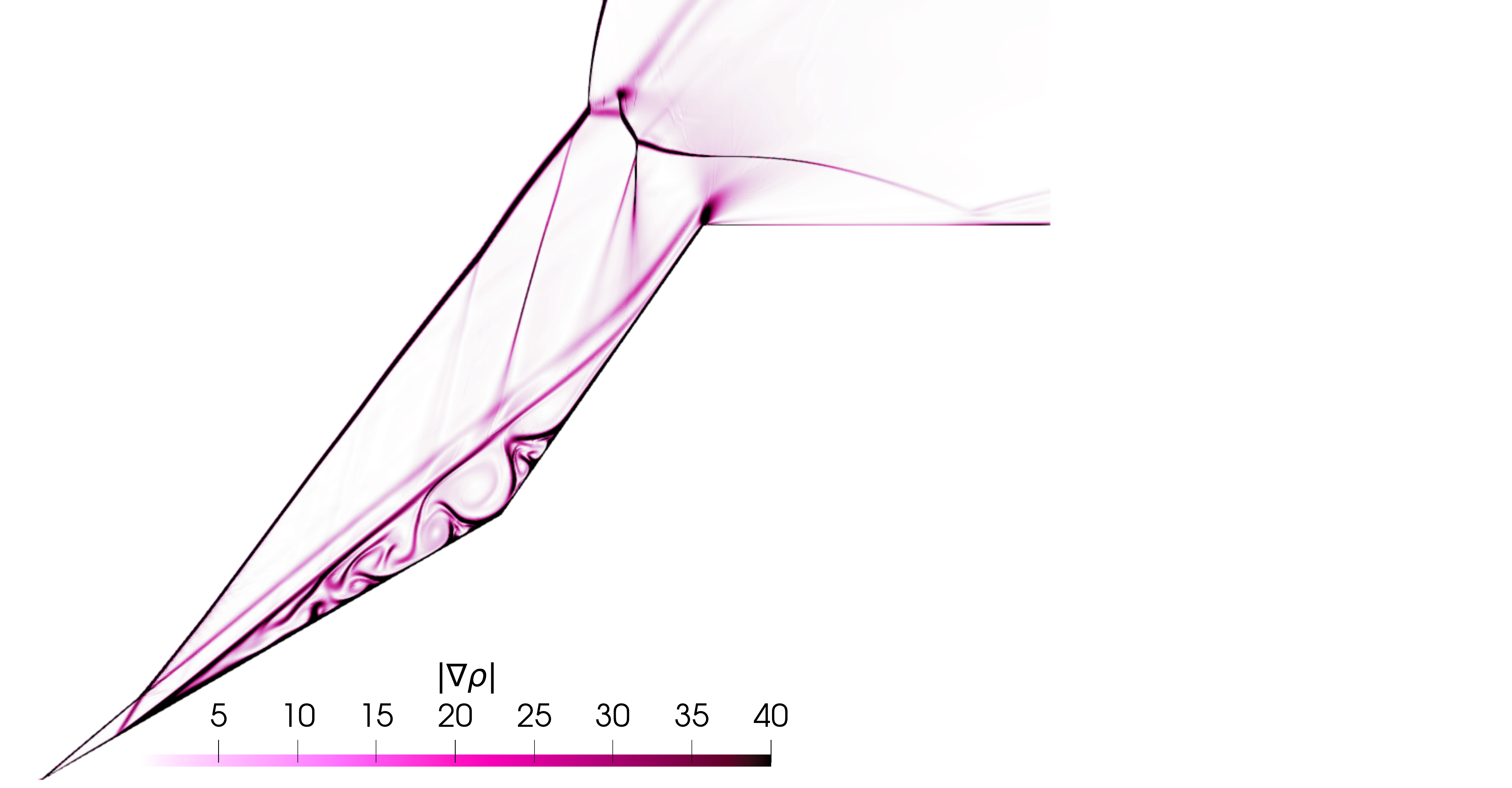}  
		\caption{}
		\label{fig:30_55_1p5_8ms}
	\end{subfigure}
	\caption{A numerical schlieren visualization of shock waves and separation region at $L_1/L_2=1.5$ and $\theta_2 = 55^\circ.$ (\subref{fig:30_55_1p5_6p3ms}) t = 6.3 ms (\subref{fig:30_55_1p5_7ms}) t = 7 ms (\subref{fig:30_55_1p5_7p5ms}) t = 7.5 ms (\subref{fig:30_55_1p5_8ms}) t = 8 ms. Watch animation (1.mp4) for more details. \protect\href{run:./1.mp4}{(Multimedia view)}}
	\label{fig:30_55_1p5}
\end{figure}

\begin{figure}[h!]
	\centering
	\includegraphics[width=\linewidth, trim={0 0 20cm 0}, clip]{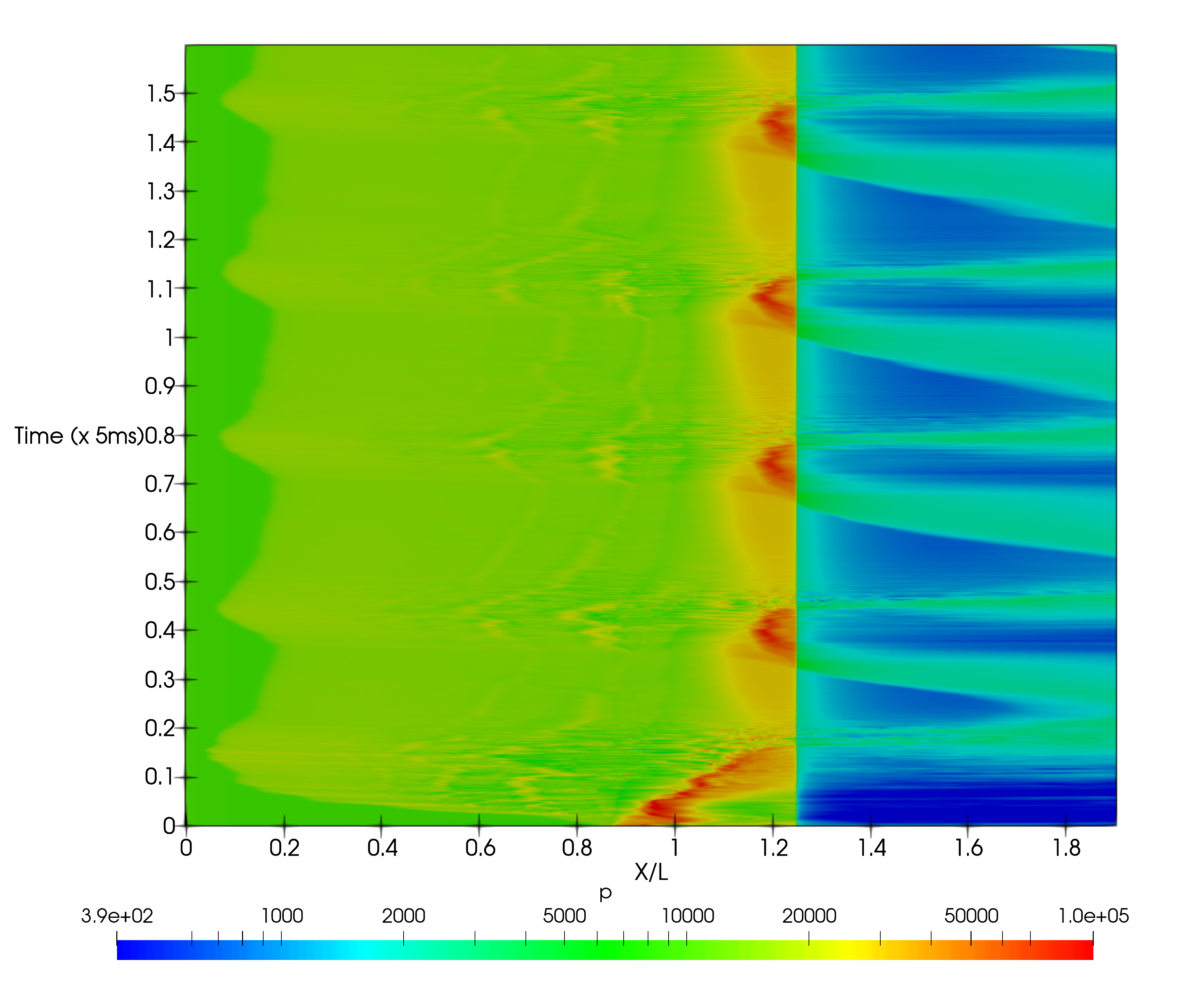}  
	\caption{A spatio-temporal variation of wall pressure at $L_1/L_2=1.5$ and $\theta_2 = 55^\circ.$}
	\label{fig:30_55_1p5_pressure}
\end{figure}

Although, for $L_1/L_2$ = 2, $\theta_1=30^\circ$ and $\theta_2=55^\circ$, it is shown that the unsteady and periodic shock-shock interaction mechanism, seen for the coarse grid, is not real because on subsequent grid refinments such oscillations disappear. But such differences are due to change in location of incidence shock as the size of separation region changes. The interaction mechanism observed with a coarse mesh encourages us to investigate it for a parameteric variation of the geometry where such interaction can be seen realistically on a refined grid. To that end, the same configuration has been simulated for $L_1/L_2$ = 1.5 such that fore-wedge length is kept constant (=50.8 mm). Due to increased length of the aft-wedge, the separation region is enlarged by a small amount, and more importantly, the expansion corner comes closer to the transmitted shock resulting in an unsteady and periodic shock-shock interaction mechanism. This can be seen in fig. \ref{fig:30_55_1p5} (multimedia view). In fig. \ref{fig:30_55_1p5_pressure}, a change in pressure distribution along with the wedge with respect to time, is shown. The distribution is very similar to one seen for coarse mesh simulation of $L_1/L_2$ = 2 case in fig. \ref{fig:30_55_2_Coarse_pressure}. The differences are seen in the separation point location, which has shifted more towards the leading edge of the fore-wedge. Also, the frequency of oscillation is larger than seen before in the case of coarse grid simulation on $L_1/L_2$ = 2 case due to growth of separation region size.

It can be seen from fig.\ref{fig:30_55_1p5_pressure} that the oscillation cycles have a time-period of approximately 1.7 ms. Also, it can be seen from fig. \ref{fig:30_55_1p5_6p3ms} and \ref{fig:30_55_1p5_8ms} that the two states are very similar to each other, which are 1.7 ms delayed in time. It should be noted that the size of the separation region is larger in fig. \ref{fig:30_55_1p5_6p3ms} at t = 6.3 ms compared to \ref{fig:30_55_1p5_7ms} at t = 7 ms even in the absence of any shock impingement on the aft-wedge. This is due to the slow relaxation of the separation region from the shock impingement on the aft-wedge from the previous cycle of the flow. Between t = 6.3 ms and 7 ms, as the separation region contracts, it brings the impingement point of the transmitted shock from the plane region downstream of the expansion corner to the aft-wedge region. At t = 7 ms, the reattachment point is still undisturbed, but as time passes, the adverse pressure gradient generated due to shock impingement reaches the reattachment region and disturbs the separation region. This causes the separation point to move upstream and changes the shock-shock interaction pattern in the flow outside the separation region. Due to a change in shock-shock interaction pattern, there is no transmitted shock impinging on the aft-wedge. This again causes relaxation of the separation region and brings the state of the flow at t = 8 ms to a state similar to one seen at t = 6.3 ms completing a cycle. This pattern is seen for $L_1/L_2 = 1.5$ and not $L_1/L_2 = 2$ mainly because of the geometric constraint. For $L_1/L_2 = 2$ case, as the separation region relaxes for the first time when the flow is still developing, the transmitted shock never crosses the expansion corner to impinge on the aft-wedge due to shorter aft-wedge length. So, the separation region relaxes to the smallest stable size possible, and a quasi-steady state is achieved.

\begin{figure}[h!]
	\centering
	\begin{subfigure}[t]{0.545\linewidth}
		\centering
		\includegraphics[width=\linewidth, trim={0 0 10cm 0}, clip]{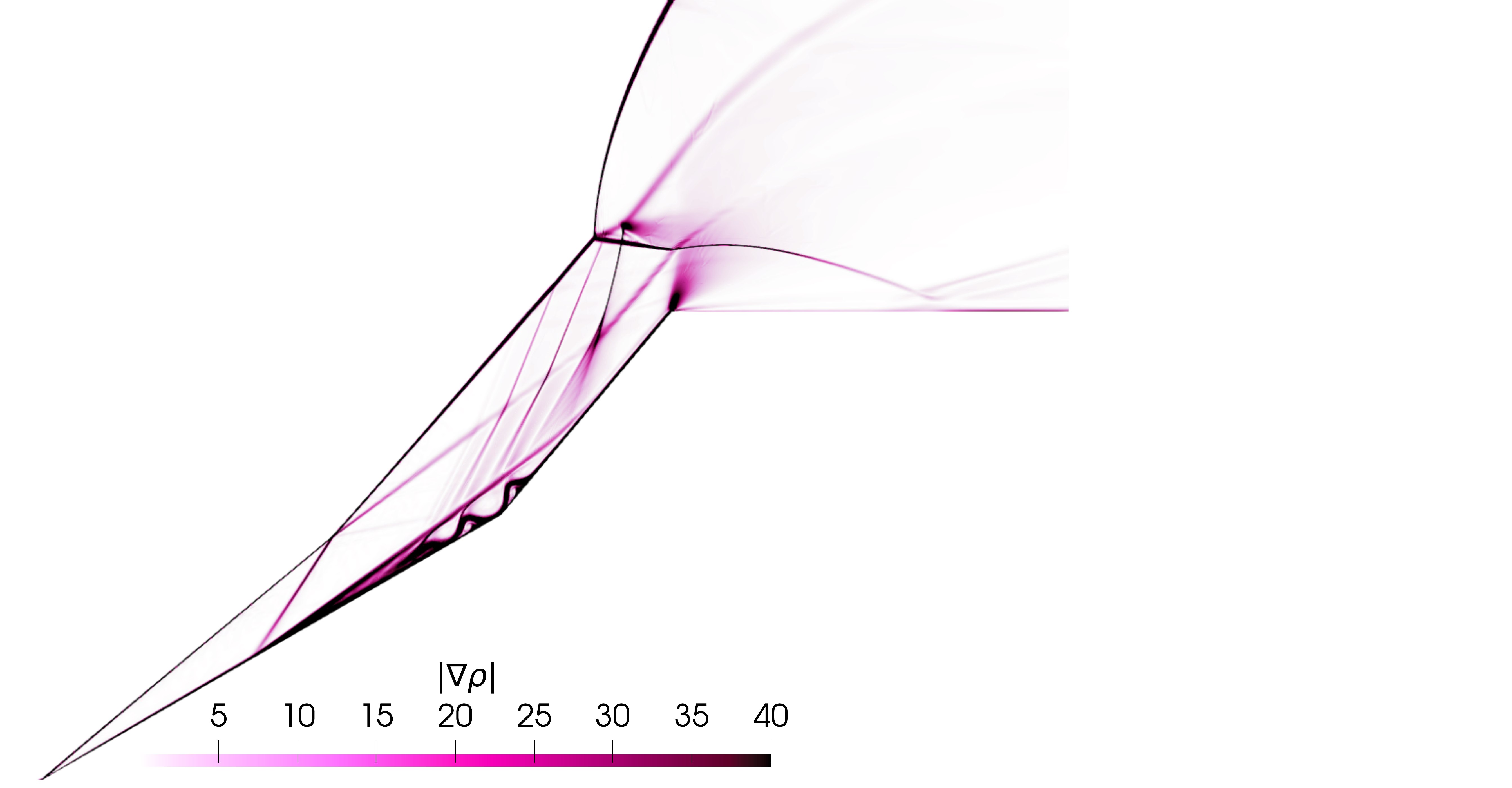}  
		\caption{}
		\label{fig:30_50_2_gradRho}
	\end{subfigure}
	\begin{subfigure}[t]{0.445\linewidth}
		\centering
		\includegraphics[width=\linewidth, trim={0 0 20cm 0}, clip]{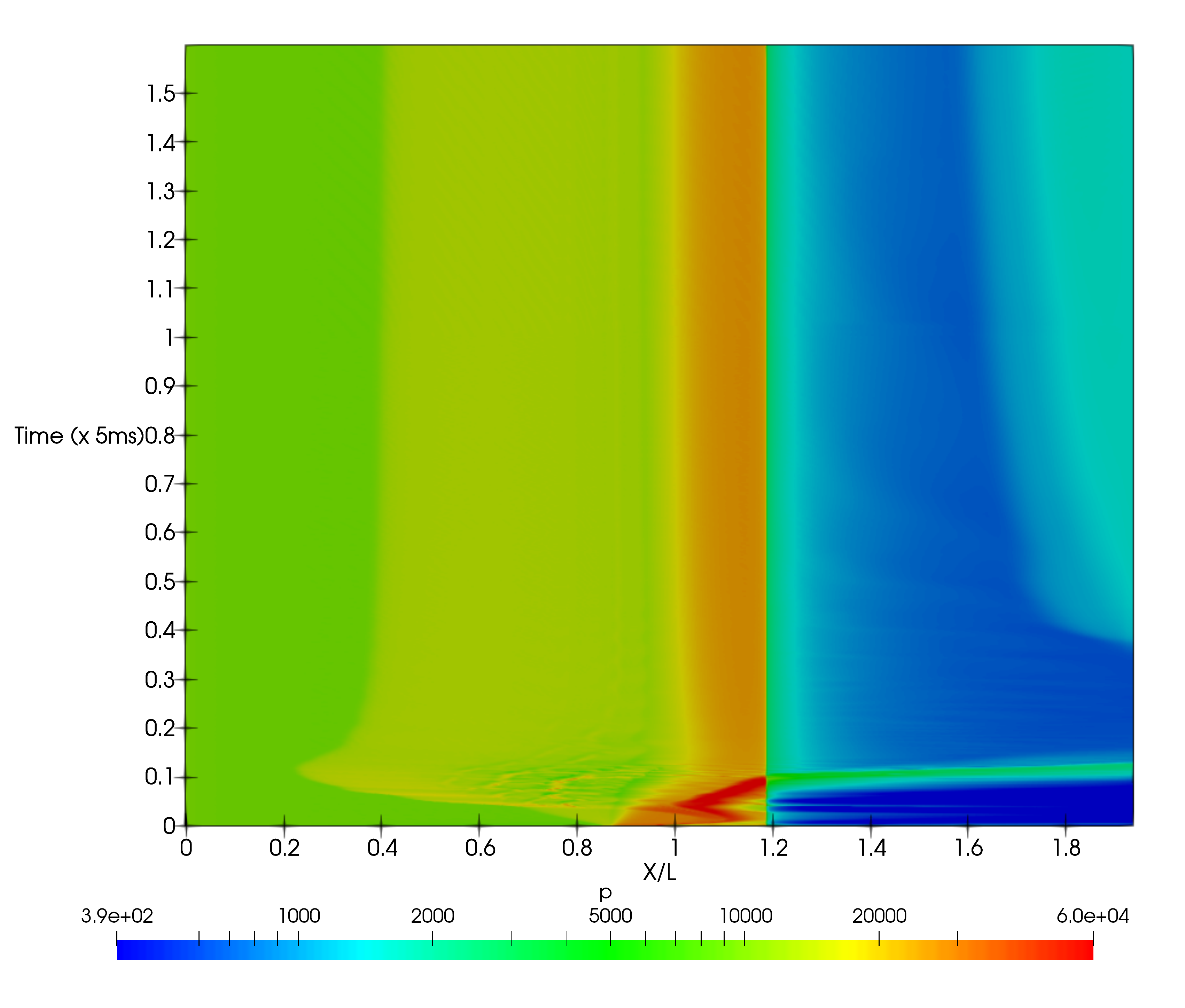}  
		\caption{}
		\label{fig:30_50_2_pressure}
	\end{subfigure}
	\caption{A numerical schlieren visualization of shock waves and separation region and spatio-temporal variation of wall pressure at $L_1/L_2=2$ and $\theta_2 = 50^\circ.$  (\subref{fig:30_50_2_gradRho}) Density gradient ($|\nabla \rho|$) (\subref{fig:30_50_2_pressure}) Wall pressure variation. Watch animation (2.mp4) for more details. \protect\href{run:./2.mp4}{(Multimedia view)}}
	\label{fig:30_50_2}
\end{figure}

This mechanism is further assessed for a smaller aft-wedge angle $\theta_2 = 50^\circ$ where the separation region will be smaller than in the previous case with $\theta_2 = 55^\circ$. The simulated density gradient and spatio-temporal variation of pressure is shown in fig. \ref{fig:30_50_2} (multimedia view), which has $L_1$ = 50.8 mm, $L_1/L_2 = 2$, $\theta_1 = 30^\circ$ and $\theta_2 = 50^\circ$ with the fine grid resolution. As expected, the size of the separation region is significantly smaller than the $\theta_2 = 55^\circ$ case, and the flow reaches a complete steady state in the absence of any tertiary vortex at the compression corner (CC).

\begin{figure}[h!]
	\centering
	\includegraphics[width=\linewidth]{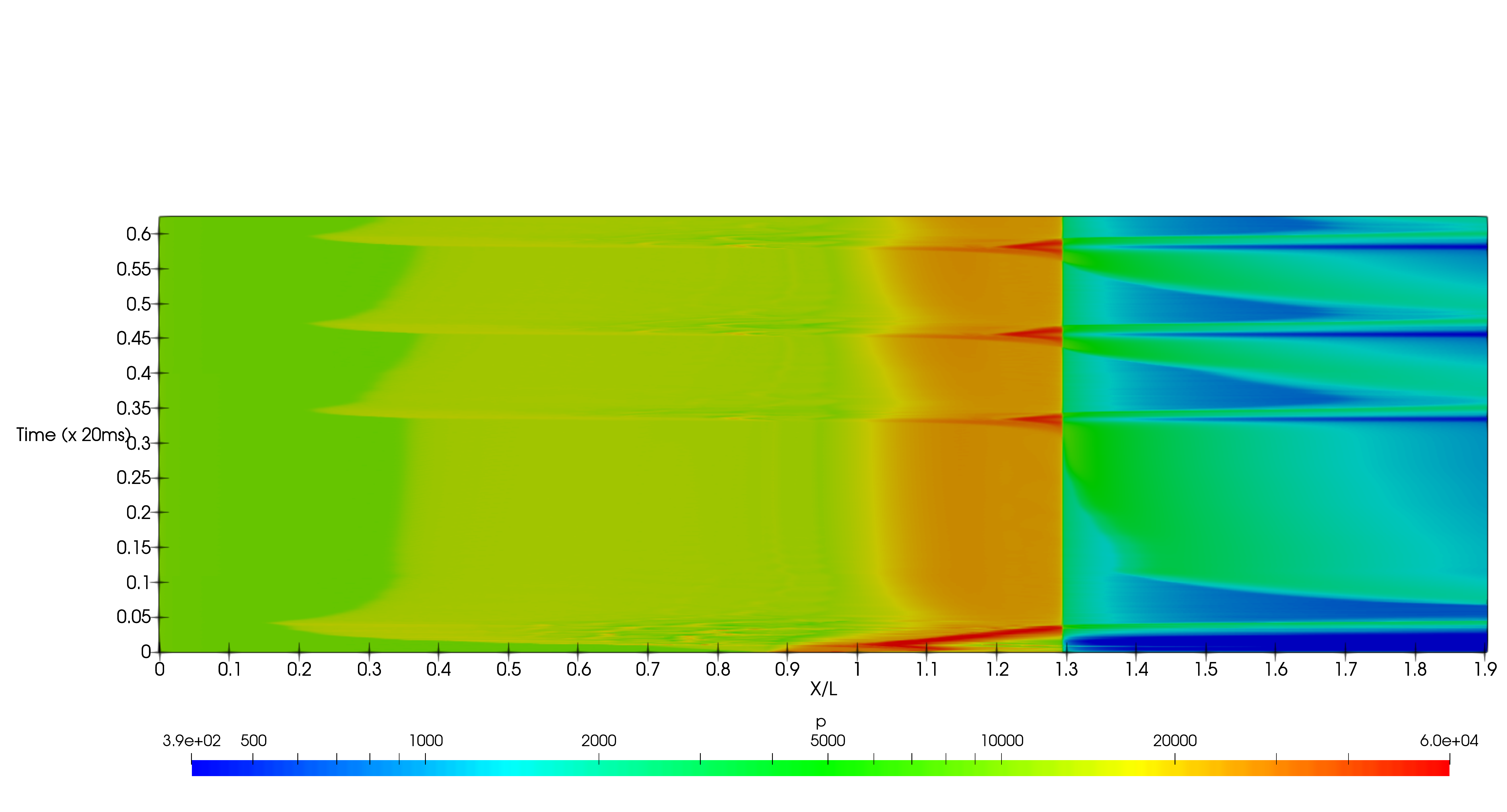}  
	\caption{A spatio-temporal variation of wall pressure at $L_1/L_2=1.5$ and $\theta_2 = 50^\circ.$}
	\label{fig:30_50_1p5_pressure}
\end{figure}

\begin{figure}[h!]
	\centering
	\begin{subfigure}[t]{0.495\linewidth}
		\centering
		\includegraphics[width=\linewidth, trim={0 0 10cm 0}, clip]{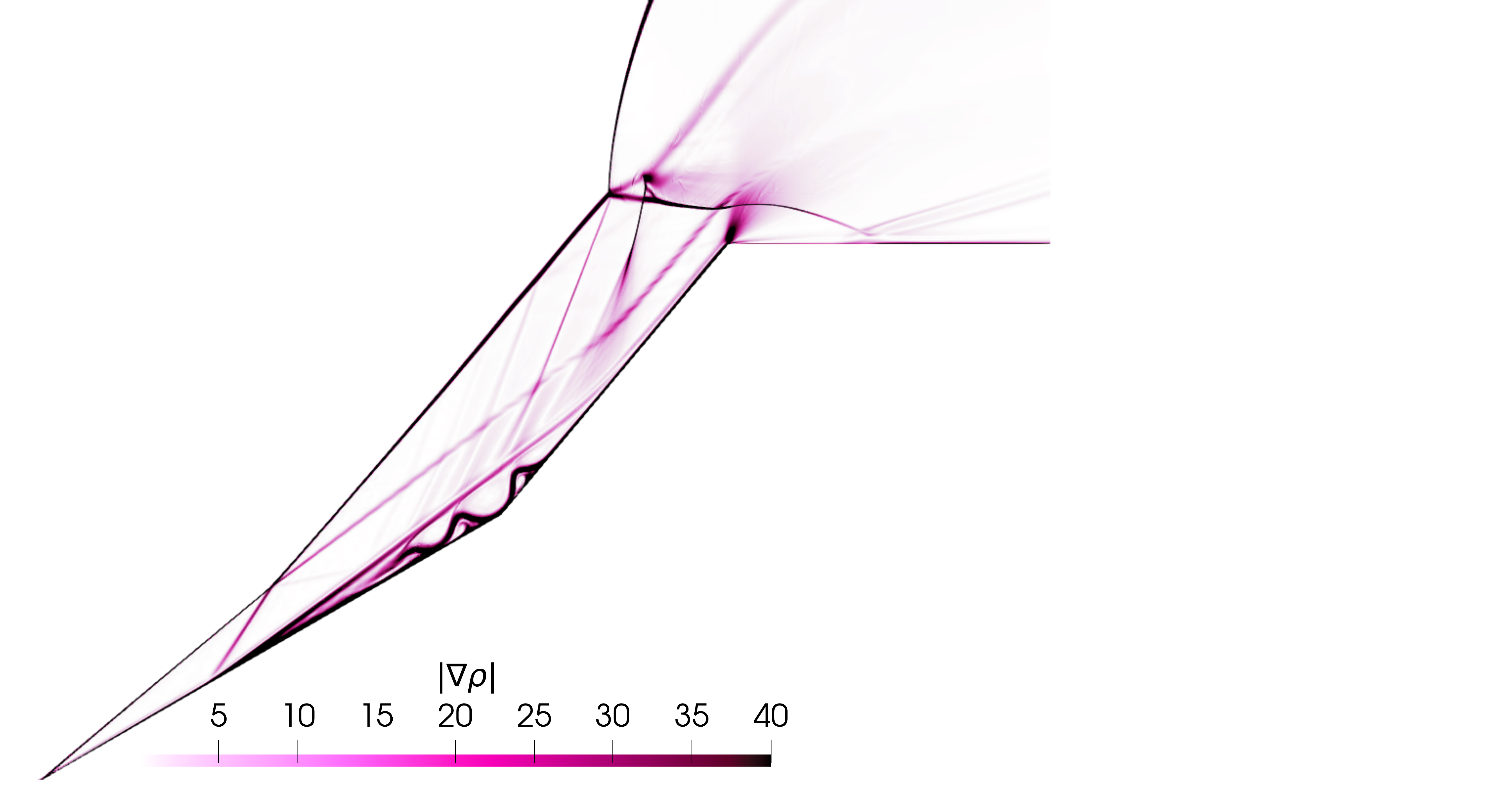}  
		\caption{}
		\label{fig:30_50_1p5_8ms}
	\end{subfigure}
	\begin{subfigure}[t]{0.495\linewidth}
		\centering
		\includegraphics[width=\linewidth, trim={0 0 10cm 0}, clip]{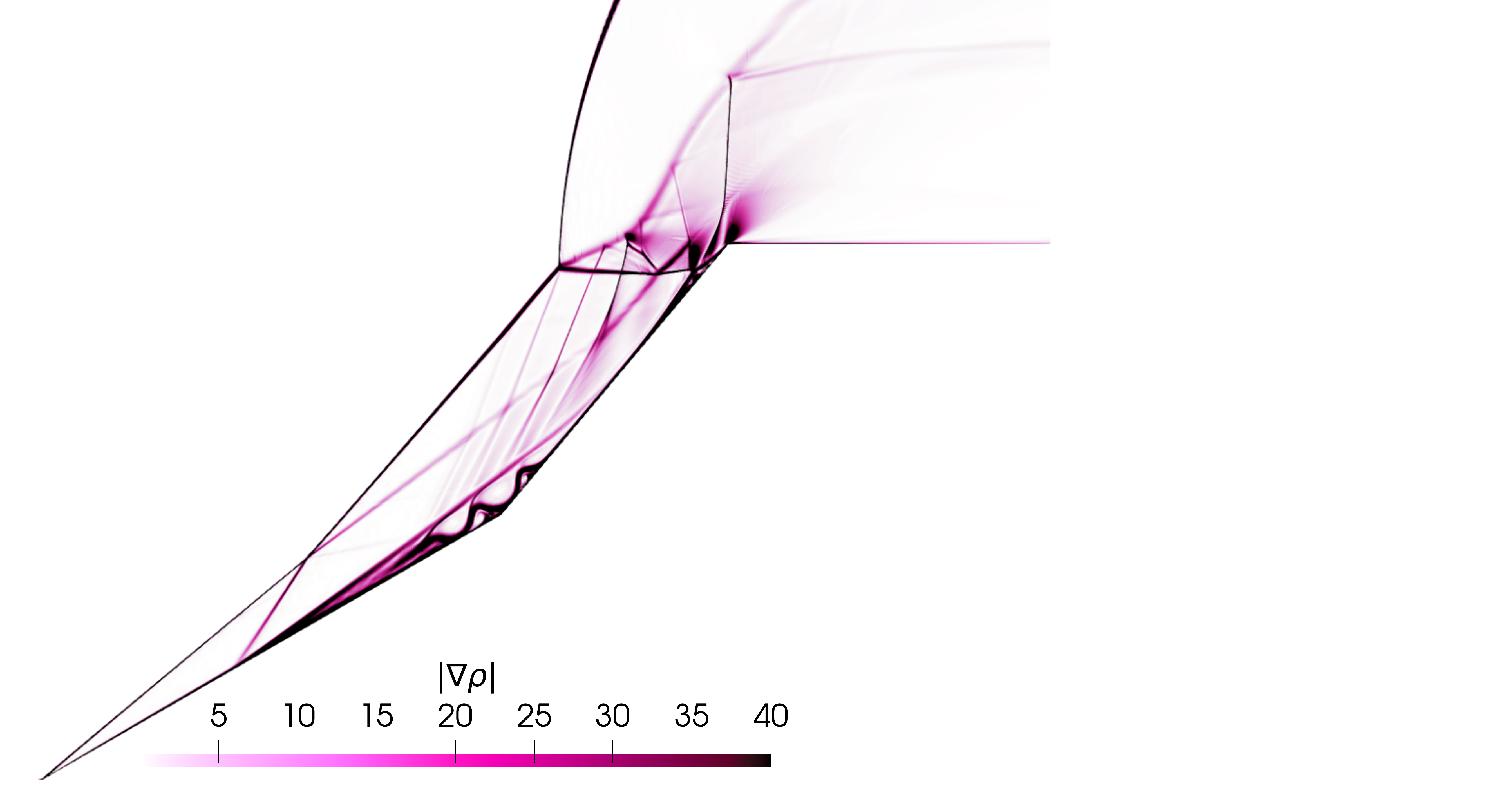}  
		\caption{}
		\label{fig:30_50_1p5_9ms}
	\end{subfigure}
	\begin{subfigure}[t]{0.495\linewidth}
		\centering
		\includegraphics[width=\linewidth, trim={0 0 10cm 0}, clip]{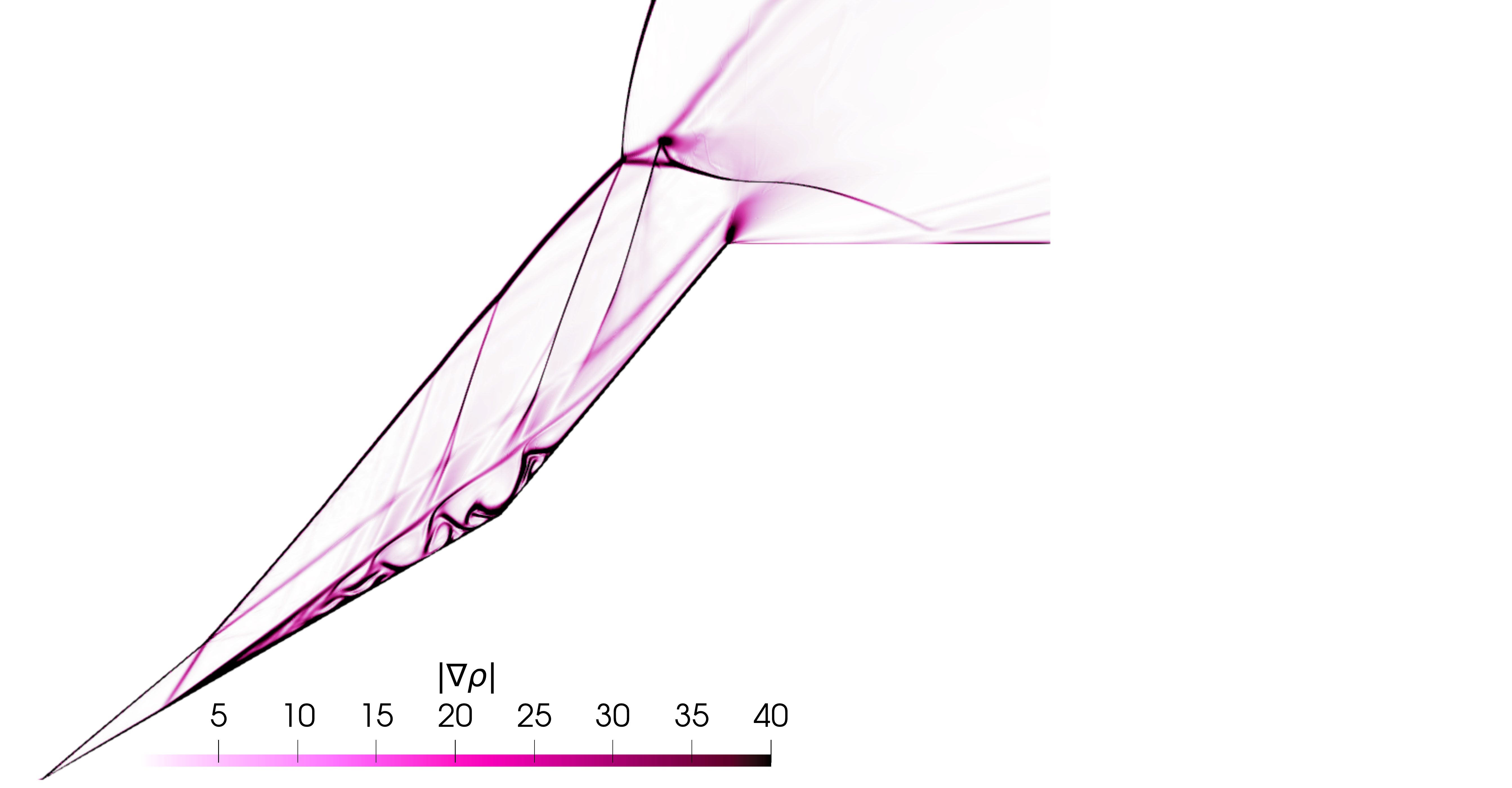}  
		\caption{}
		\label{fig:30_50_1p5_9p5ms}
	\end{subfigure}
	\begin{subfigure}[t]{0.495\linewidth}
		\centering
		\includegraphics[width=\linewidth, trim={0 0 10cm 0}, clip]{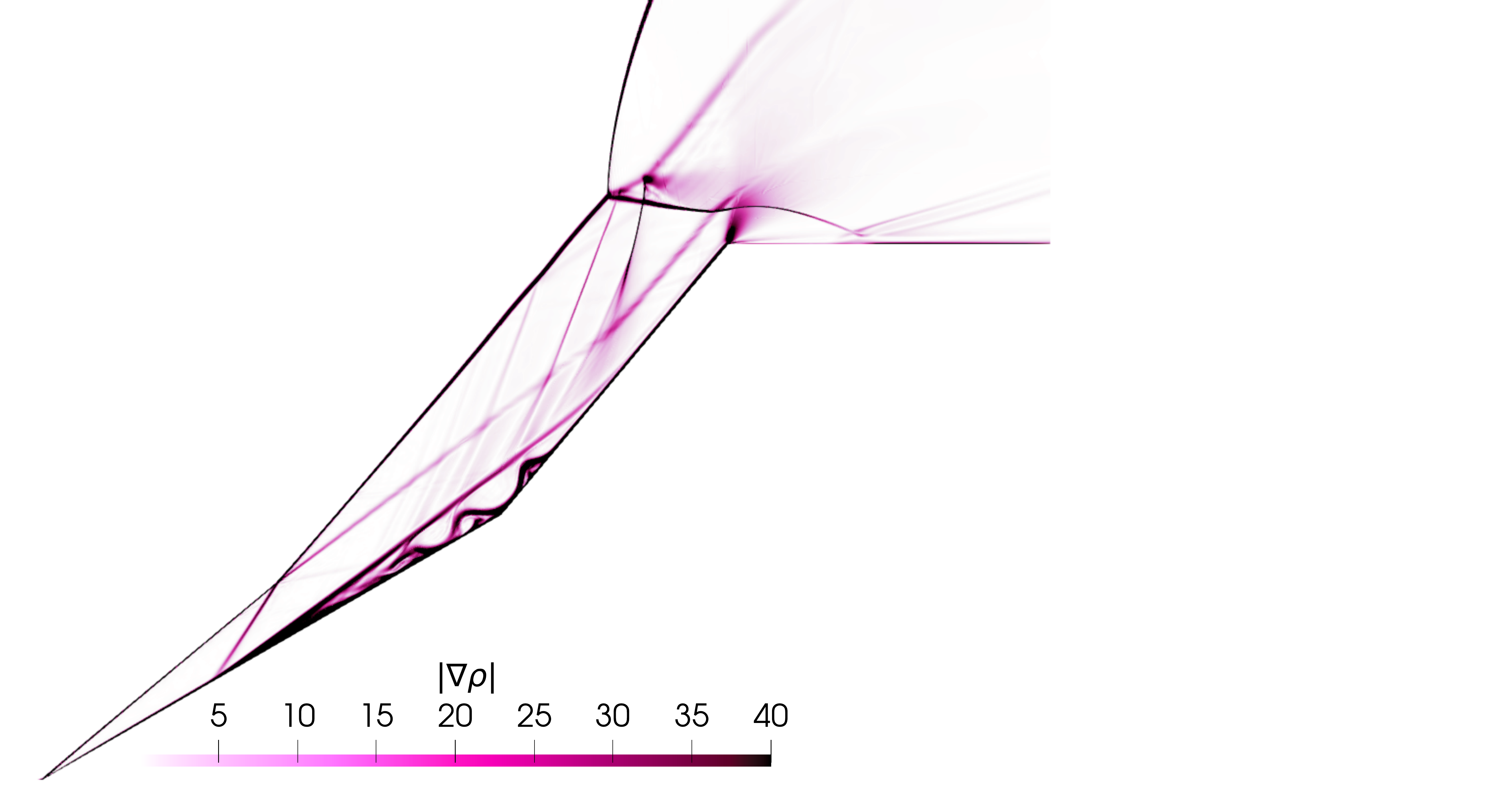}  
		\caption{}
		\label{fig:30_50_1p5_10p25ms}
	\end{subfigure}
	\caption{A numerical schlieren visualization of shock waves and separation region at $L_1/L_2=1.5$ and $\theta_2 = 50^\circ.$ (\subref{fig:30_50_1p5_8ms}) t = 8 ms (\subref{fig:30_50_1p5_9ms}) t = 9 ms (\subref{fig:30_50_1p5_9p5ms}) t = 9.5 ms (\subref{fig:30_50_1p5_10p25ms}) t = 10.25 ms. Watch animation (3.mp4) for more details. \protect\href{run:./3.mp4}{(Multimedia view)}}
	\label{fig:30_50_1p5}
\end{figure}

Figure \ref{fig:30_50_1p5_pressure} shows the spatio-temporal variation of pressure over the wedge for $\theta_2 = 50^\circ$ and $L_1/L_2 = 1.5$. It can be clearly seen that the flow achieves a time-periodic oscillatory state. It should also be noted that it takes a significantly long time to reach this periodic state of flow as compared to $\theta_2 = 55^\circ$ case. This difference should be attributed to the slower relaxation of the separation region at a smaller aft-wedge angle. From fig. \ref{fig:30_50_1p5_pressure} as well as fig.\ref{fig:30_50_1p5} (multimedia view), it can be seen that the period of oscillation is approximately 2.25 ms as the flow states in fig.\ref{fig:30_50_1p5_8ms} and \ref{fig:30_50_1p5_10p25ms} are almost identical. In fig. \ref{fig:30_50_1p5_9ms}, the separation region is comparatively smaller than the one seen at t = 8 ms, and the incidence shock impinges on the aft wedge. This disturbance near the reattachment point causes the separation region to grow and pushes the incidence shock away from the aft-wedge. This results in a further relaxation of the separation region, and the cycle repeats. 

\section{Conclusion}
A more accurate numerical simulation is performed for a hypersonic flow over a double wedge configuration. This is achieved using a more stable numerical solver as well as a better-resolved grid. A systematic grid comparison study is performed to establish the true nature of the flow. It is shown that with proper accuracy of the solution a quasi-steady flow is achieved for double wedge configuration at $\theta_1=30^\circ$, $\theta_2=55^\circ$, and $L_1/L_2=2$ at Mach number = 7. This is in contradiction to the general consensus among the published literature that after a long time, the flow settles down into a low frequecy periodically oscillating state for this configuration. It is also shown that by changing the wedge length ratio, a geometric constraint can be achieved where overall flow settles down in an oscillatory state, similar to one seen in previous literature. It is shown that the size of the separation region influences the overall shock-shock interaction pattern, which finally influences the steady or unsteady nature of the flow. The location of the incidence shock impingement point is more important while determining the nature of the developed flow, as seen by a comparison of $\theta_2 = 50^\circ$ and $55^\circ$ cases. Hence, the presence of incidence shock just downstream of the expansion corner while the separation region is overstretched is the single important parameter determining if the flow is going to be steady or periodic in nature. Different configuration can be achieved by varying different geometric parameters such as wedge lengths and wedge angles but a periodic state is achieved only when the incidence shock crosses the expansion corner to impinge on the aft-wedge while the over-stretched separation region is relaxing. This disturbes the process of separation region relaxation and makes the flow repeat the previous cycle.


\section*{Data Availability}
The data that support the findings of this study are available from the corresponding author upon reasonable request.

\section*{Supplementary Material}
See \protect\href{run:./4.mp4}{video (4.mp4)} for animation of the flow simulation with $L_1/L_2=2$ and $\theta_2=55^\circ$

\begin{acknowledgments}
Simulations are carried out on the computers provided by National PARAM Supercomputing Facility (CDAC) (www.cdac.in), and the manuscript preparation, as well as data analysis, has been carried out using the resources available at IITK (www.iitk.ac.in/cc). The support is gratefully acknowledged.
\end{acknowledgments}

\bibliography{aipsamp}

\appendix
\section{Solver Validation\label{Validation}}
The improved {\it rhoCentralFoam} solver has been extensively tested in various flow regimes. For the brevity of this manuscript and to validate the performance of the improved {\it rhoCentralFoam} solver in the context of high speed flows, three cases, namely a supersonic jet, supersonic flow over a prism, and hypersonic flow past a circular cylinder, is considered here.

\subsection{\label{subsec:jet}Supersonic jet}
\begin{figure}[h!]
	\centering
	\includegraphics[width=\textwidth]{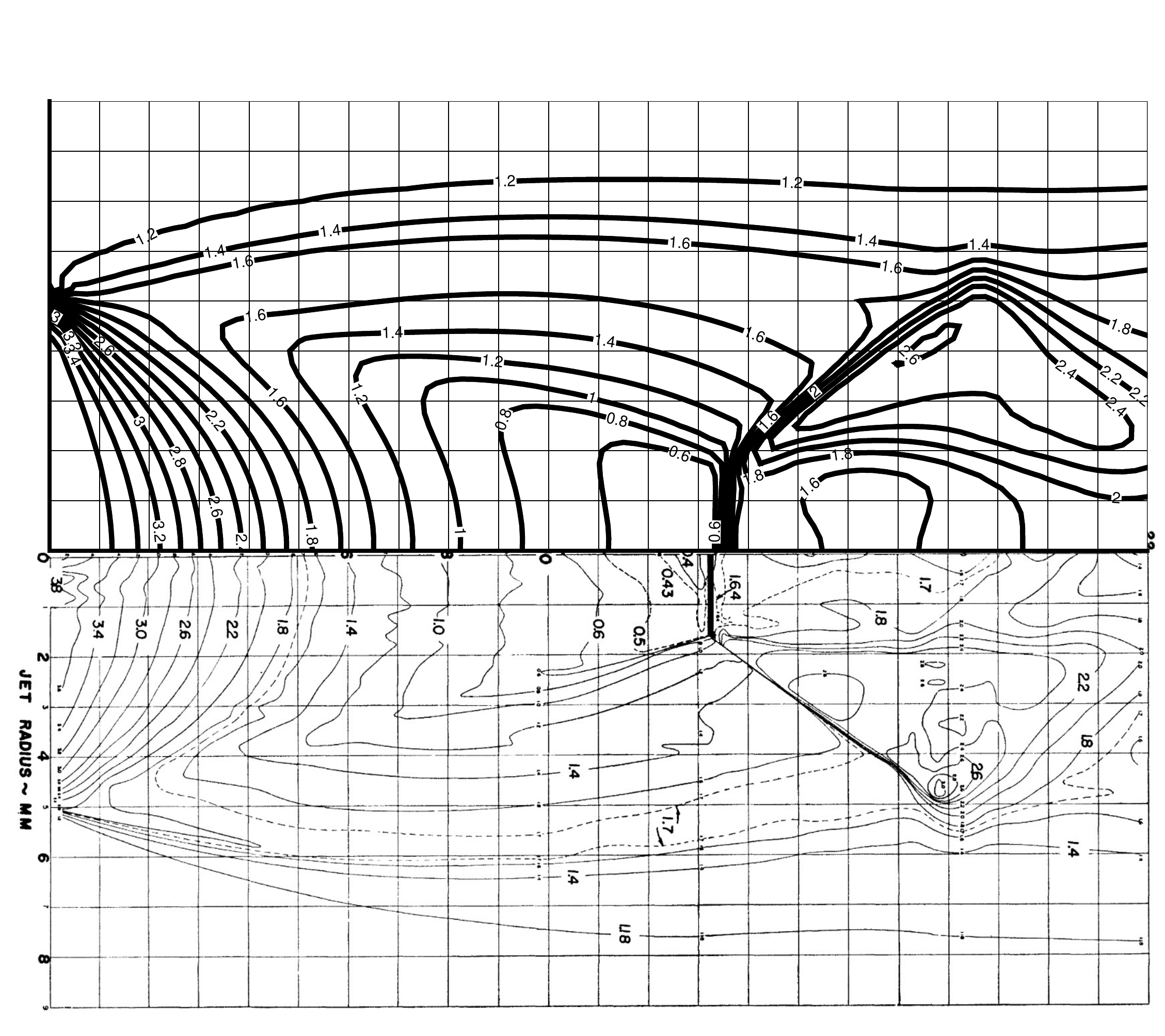}
	\caption{Comparison of density contours in Ludenburg jet. Our simulation (top) and original experimental data (bottom)}
	\label{fig:jetRho_contour}
\end{figure}

In this subsection, a 2D axisymmetric supersonic free jet flow has been validated against the experiments of Ladenburg et al. \cite{ladenburg1949interferometric}. A case with the tank pressure 60 $lb/in^2$ or 4.14 $bar$ is chosen based on recommendations of ref. \cite{greenshields2010implementation}, which suggests that it produces a Mach stem that is challenging to reproduce, accounting the correct amount of viscous and numerical dissipation. An exactly the same case setup is used in the current simulation as followed in ref. \cite{greenshields2010implementation}, and an excellent match of density contours with experimental data of ref. \cite{ladenburg1949interferometric} is seen without any spurious oscillations. The simulation is run using the TVD RK-3 time integration method at CFL 1.0. Figure \ref{fig:jetRho_contour} shows the density contours comparison of simulation using the new solver with the experiment\cite{ladenburg1949interferometric}, and an excellent match in the location of the Mach stem shows the small numerical dissipation for high Mach number flows.

\subsection{\label{subsec:prism}Supersonic flow over a prism}
For estimation of numerical stability and numerical dissipation of an inviscid as well as viscous solver, the shock-vortex interaction in schardin's problem studied by Chang et al. in ref. \cite{chang2000shock} is adopted here. In this problem, a shock wave is traversing through an equilateral triangle (or wedge) at shock Mach number $M_s = 1.34$. The side of the triangular wedge is 20 mm and is confined within a shock tube of width 150mm while the wedge being placed at the centerline. Time t=0 is assigned at the event when the shock wave hits the leading edge of the wedge for the first time. 

\begin{figure}[h!]
	\centering
	\includegraphics[width=0.8\linewidth]{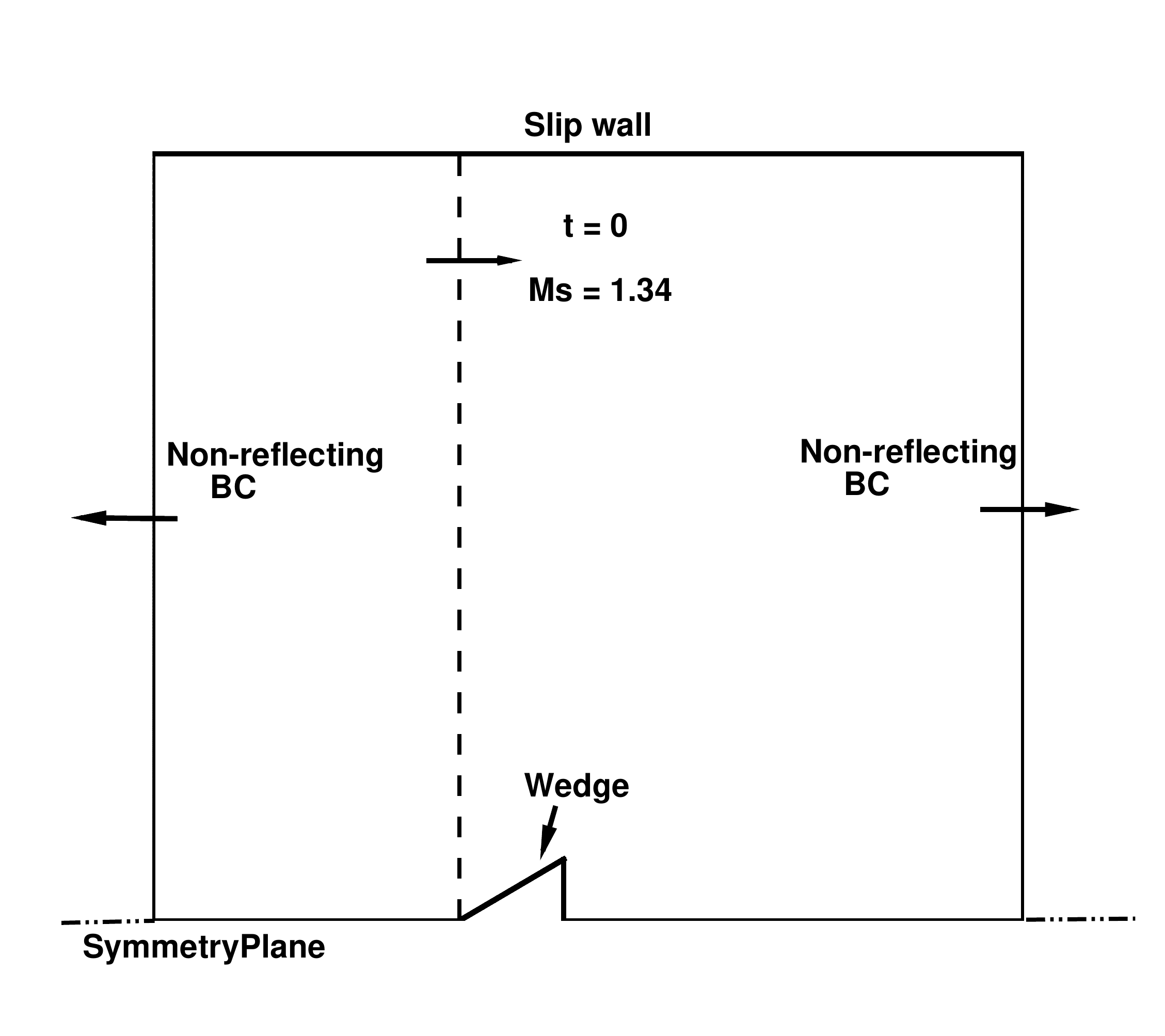}  
	\caption{Domain and boundary conditions for shock-vortex interaction flow}
	\label{fig:wedgeDomain}
\end{figure}

\begin{figure}[h!]
	\centering
	\includegraphics[width=\linewidth] {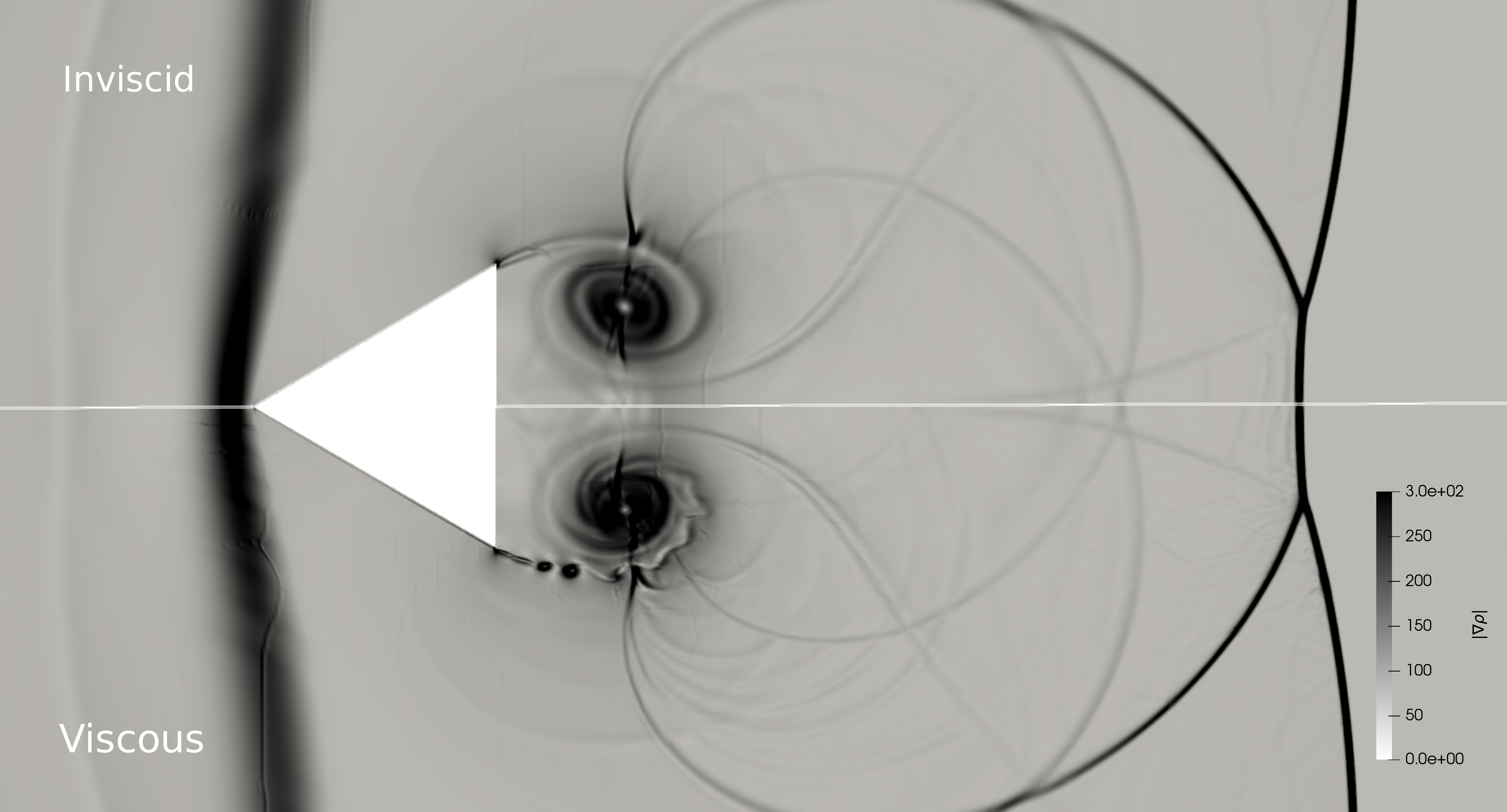}  
	\caption{Density gradient field from computational results at time = 175$\mu s$. t = 0 is measured from the time shock wave first arrives at the leading edge of the wedge}
	\label{fig:prismgradRho}
\end{figure}

\begin{figure}[h!]
	\centering
	\includegraphics[width=\linewidth]{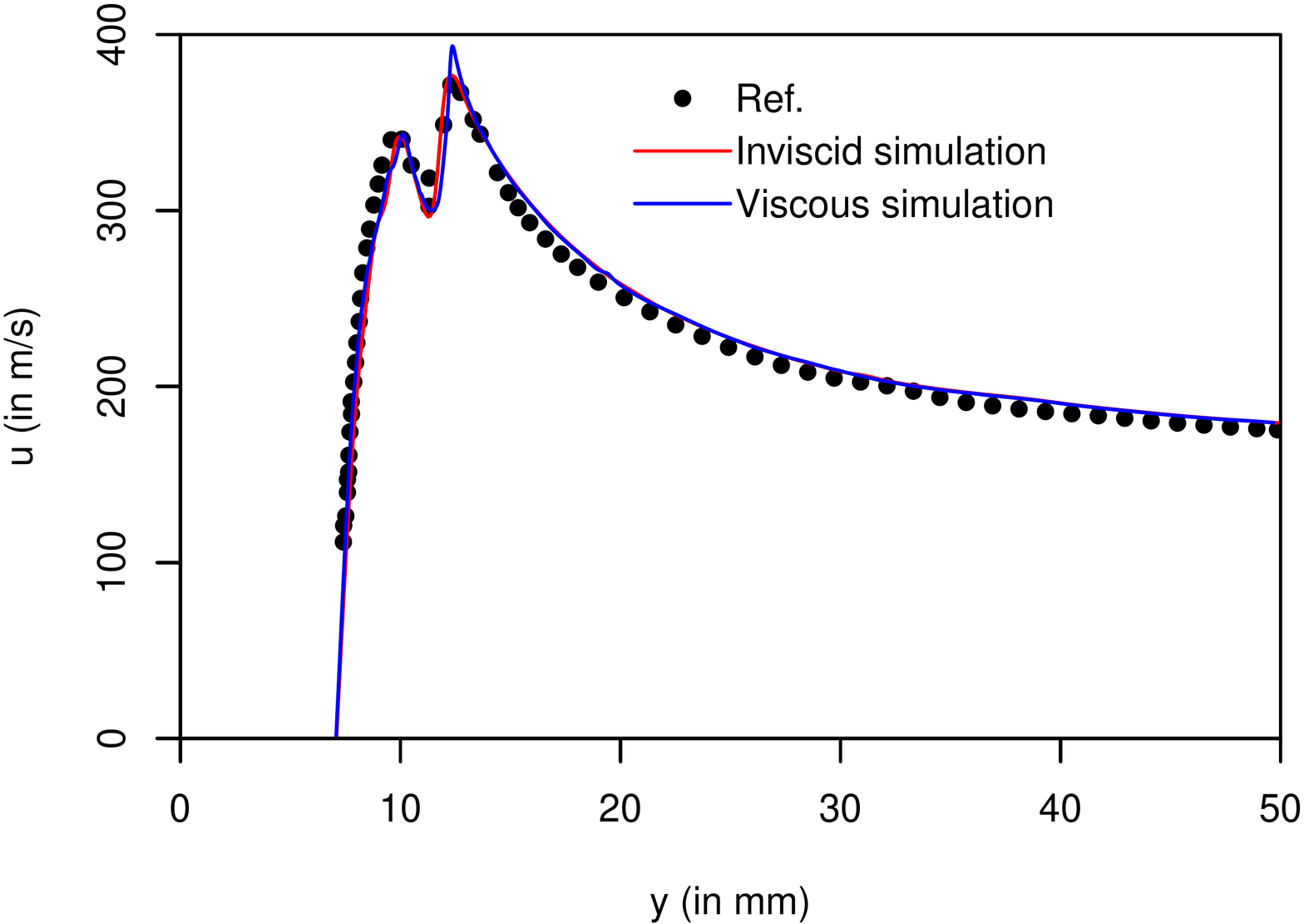}  
	\caption{Comparison of the computed tangential velocity distribution of primary vortex with computational results of ref.\cite{chang2000shock}}
	\label{fig:SVIuy}
\end{figure}

As shown in fig. \ref{fig:wedgeDomain}, the same computational domain is adopted as in ref. \cite{chang2000shock} with slip velocity conditions at the wedge, top, and the bottom wall and non-reflecting boundary conditions at the inlet and the outlet. For viscous simulation, the wedge is assigned a no-slip adiabatic boundary condition. Simulation is performed on a grid where cell size is of the order 0.01 mm behind the wedge and 0.1 mm in the far-field. As shown in fig. \ref{fig:prismgradRho}, this solver can predict an accurate inviscid as well as viscous features of shock vortex interactions and shear layer instability as mentioned in ref. \cite{chang2000shock,seshadri2020investigation} without any oscillation or dissipation of structures using the TVD RK-3 method at CFL 1.0. The new solution method and the use of a higher-order time discretization scheme have improved the performance of the solver to a greater extent. An excellent match is obtained for comparison of computed tangential velocity in the primary vortex region with the published results of ref. \cite{chang2000shock} in fig. \ref{fig:SVIuy}. 

\subsection{\label{subsec:Cylinder}Hypersonic flow past a circular cylinder}
\begin{figure}[h!]
	\centering
	\begin{subfigure}{\linewidth}
		\centering
		\includegraphics[width=\linewidth]{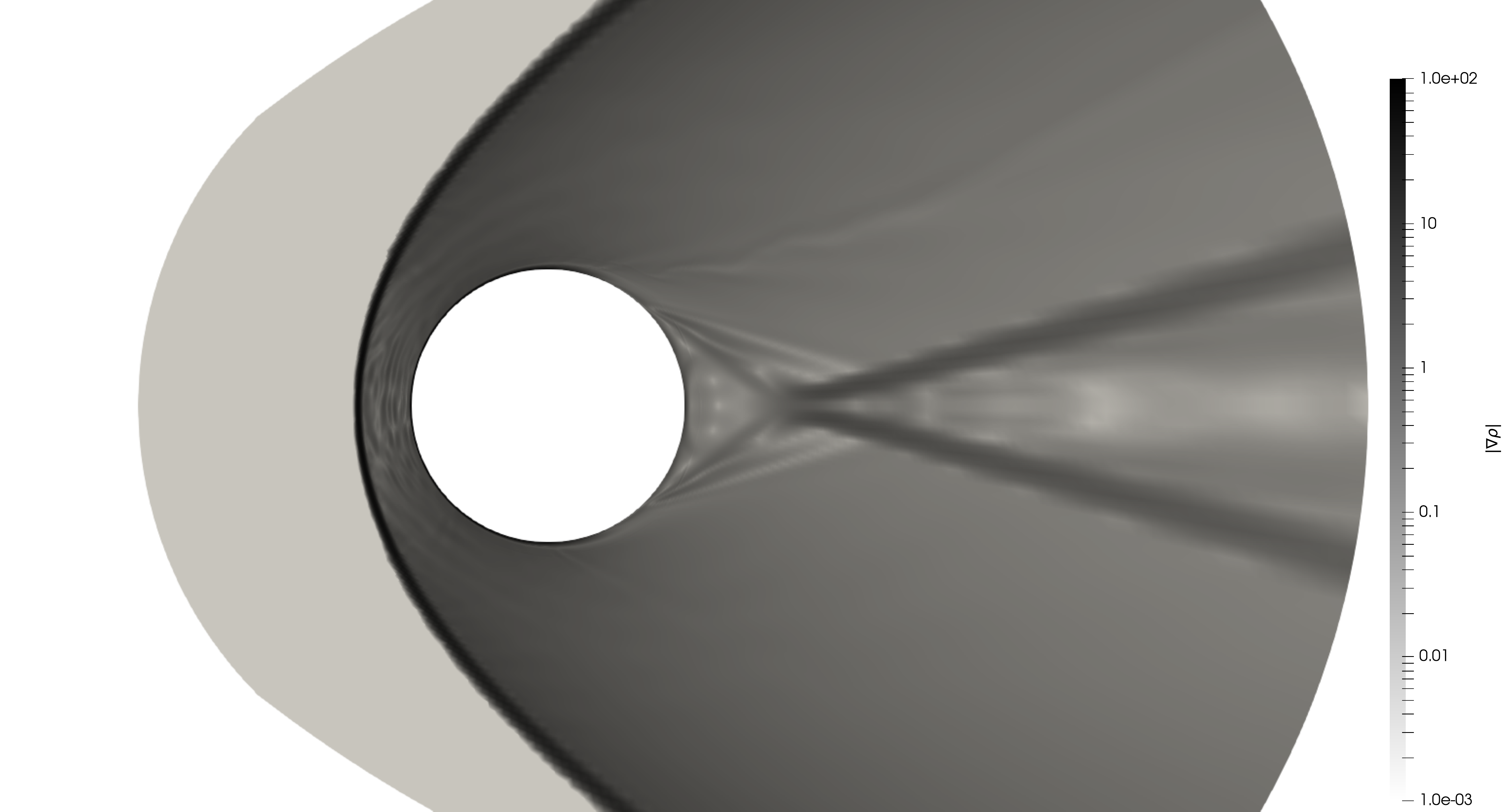}  
		\caption{}
		\label{fig:cylinderGradRho}
	\end{subfigure}
	\begin{subfigure}{0.8\linewidth}
		\centering
		\includegraphics[width=\linewidth]{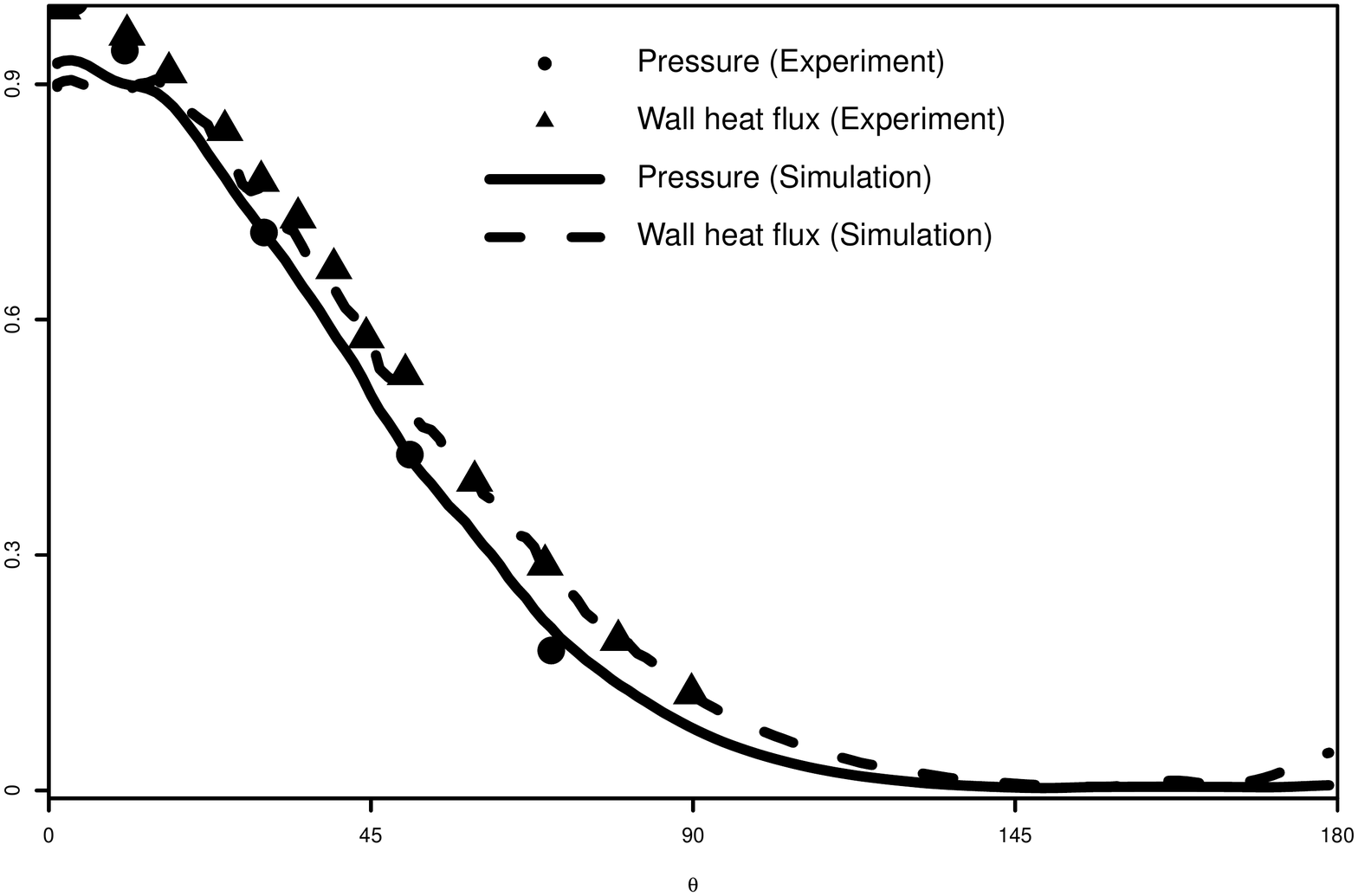}  
		\caption{}
		\label{fig:cylinderpq}
	\end{subfigure}
	\caption{Computational results for hypersonic flow over a circular cylinder. (\subref{fig:cylinderGradRho}) Density gradient ($|\nabla\rho|$) (\subref{fig:cylinderpq}) Comparison of surface pressure and wall heat flux with experimental results of ref.\cite{wieting1989experimental}}
	\label{fig:gradRho}
\end{figure}

For validation of 2D hypersonic viscous flow, a case with a rather complex flow physics of flow past a cylinder is chosen. In this problem, the flow field is full of complex features, albeit the geometric complexity is next to none. Simulation is performed using a structured grid with 300 divisions in the azimuthal direction and 225 divisions in the radial direction. Details of flow parameters are given in Table \ref{tab:cylinder_thermo_properties}. Flow features around the cylinder such as detached bow shock in the front and a recirculation region in the cylinder wake are seen in fig. \ref{fig:cylinderGradRho} to be very accurately predicted qualitatively. For quantitative measurements, normalized wall pressure($p/p_0$) and normalized wall heat flux rate ($q/q_0$) are shown in fig. \ref{fig:cylinderpq} compared to the experimental results of ref.\cite{wieting1989experimental}. The prediction of wall heat flux for hypersonic flow is a challenging task which is very accurately predicted by this flow solver. Hence, we can safely assume the current flow solver can be reliably used for unsteady simulations and wall heat flux prediction of flow over a double wedge for the long time durations.   
\begin{table}[h!]
	\centering
	\begin{tabular}{|l | l |}\hline
		Parameters & value \\ \hline
		Pressure $(p_\infty)$ & 855 Pa\\
		Temperature $(T_\infty)$ & 201 K\\
		Mach No. $(M_\infty) $ & 6.5\\
		Reynolds No. $(Re_\infty)$ & $1.55\times10^5$\\
		cylinder diameter (D) & 7.62 cm\\ \hline
	\end{tabular}
	\captionof{table}{Geometric and flow parameters for hypersonic cylinder flow}
	\label{tab:cylinder_thermo_properties}
\end{table}
\end{document}